\documentclass[sigconf]{acmart}

\usepackage{enumitem}
\usepackage{multirow}
\usepackage{lipsum}
\usepackage{listings}
\usepackage{subfigure}
\usepackage{algorithmic}
\usepackage{graphicx}
\usepackage{pdfpages}
\usepackage{verbatim}
\usepackage{pifont}
\usepackage{etoolbox}
\usepackage{textcomp}
\usepackage{ulem}
\usepackage{xcolor}
\usepackage{hyperref}
\usepackage{url}
\usepackage{caption}

\AtBeginDocument{%
  \providecommand\BibTeX{{%
    \normalfont B\kern-0.5em{\scshape i\kern-0.25em b}\kern-0.8em\TeX}}}

\setcopyright{acmcopyright}
\acmYear{2021}\copyrightyear{2021}
\setcopyright{rightsretained}
\acmConference[ICS '21]{2021 International Conference on Supercomputing}{June 14--17, 2021}{Virtual Event, USA}
\acmBooktitle{2021 International Conference on Supercomputing (ICS '21), June 14--17, 2021, Virtual Event, USA}
\acmPrice{}
\acmDOI{10.1145/3447818.3460364}
\acmISBN{978-1-4503-8335-6/21/06}

\begin{document}

\title{FT-BLAS: A High Performance BLAS Implementation With Online Fault Tolerance}

\author{Yujia Zhai}
\affiliation{%
  \institution{University of California, Riverside}
  \city{Riverside}
  \state{CA}
  \country{USA}}
\email{yzhai015@ucr.edu}

\author{Elisabeth Giem}
\affiliation{%
  \institution{University of California, Riverside}
  \city{Riverside}
  \state{CA}
  \country{USA}}
\email{gieme01@ucr.edu}

\author{Quan Fan}
\affiliation{%
  \institution{University of California, Riverside}
  \city{Riverside}
  \state{CA}
  \country{USA}}
\email{qfan005@ucr.edu}

\author{Kai Zhao}
\affiliation{%
  \institution{University of California, Riverside}
  \city{Riverside}
  \state{CA}
  \country{USA}}
\email{kzhao016@ucr.edu}

\author{Jinyang Liu}
\affiliation{%
  \institution{University of California, Riverside}
  \city{Riverside}
  \state{CA}
  \country{USA}}
\email{jliu447@ucr.edu}

\author{Zizhong Chen}
\affiliation{%
  \institution{University of California, Riverside}
  \city{Riverside}
  \state{CA}
  \country{USA}}
\email{chen@cs.ucr.edu}


\begin{abstract}
    Basic Linear Algebra Subprograms (BLAS) is a core library in scientific computing and machine learning. This paper presents FT-BLAS, a new implementation of BLAS routines that not only tolerates soft errors on the fly, but also provides comparable performance to modern state-of-the-art BLAS libraries on widely-used processors such as Intel Skylake and Cascade Lake. To accommodate the features of BLAS, which contains both memory-bound and computing-bound routines, we propose a hybrid strategy to incorporate fault tolerance into our brand-new BLAS implementation: duplicating computing instructions for memory-bound Level-1 and Level-2 BLAS routines and incorporating an Algorithm-Based Fault Tolerance mechanism for computing-bound Level-3 BLAS routines. Our high performance and low overhead are obtained from delicate assembly-level optimization and a kernel-fusion approach to the computing kernels. Experimental results demonstrate that FT-BLAS offers high reliability and high performance -- faster than Intel MKL, OpenBLAS, and BLIS by up to 3.50\%, 22.14\% and 21.70\%, respectively, for routines spanning all three levels of BLAS we benchmarked, even under hundreds of errors injected per minute.
\end{abstract}


\maketitle

\section{Introduction}
Processor chips are more susceptible to transient faults than ever before due to common performance-enhancing technologies such as shrinking transistor width, higher circuit density, and lower near-threshold voltage operations \cite{laprie1985dependable, lutz1993analyzing, nicolaidis1999time}. Transient faults can alter a signal transfer or corrupt the bits within stored values instead of causing permanent physical damage \cite{gomez2015detecting, li2012classifying}. As a consequence, reliability has been identified by U.S. Department of Energy officials as one of the major challenges for exascale computing \cite{lucas2014doe}. 

Since Intel observed the first transient error and resulting soft data corruption in 1978 \cite{may1979alpha}, transient faults have had a significant impact on both academia and industry in the following years. Sun Microsystems reported in 2000 that server crashes caused by cosmic ray strikes on unprotected caches were responsible for the outages of random customer sites including America Online, eBay, and others \cite{baumann2002soft}. In 2003, Virginia Tech demolished the newly-built Big Mac cluster of 1100 Apple Power Mac G5 computers into individual components and sold them online because the cluster was not protected by error correcting code (ECC) and fell prey to cosmic ray-induced partial strikes, causing repeated crashes and rendering it unusable \cite{geist2016supercomputing}. Transient faults can still threaten system reliability even if a cluster is protected by ECC: Oliveira et al. simulated an exascale machine with 190,000 cutting-edge Xeon Phi processors that could still experience daily transient errors under ECC protection \cite{oliveira2017experimental}.

If an affected application crashes when a transient fault occurs, we call it a fail-stop error. If the affected application continues but produces incorrect results, we call it a fail-continue error. Fail-stop errors can often be protected by checkpoint/restart mechanisms (C/R)~\cite{phillips2005scalable, NEURIPS2019_9015, tao2018improving, tensorflow2015-whitepaper} and algorithmic approaches \cite{hakkarinen2014fail, chen2008scalable, chen2008extending}. Fail-continue errors are often more dangerous because they can corrupt application states without any warning from the system, and lead to incorrect computing results \cite{mitra2014resilience, cher2014understanding, dongarra2011international, calhoun2017towards, snir2014addressing}, which can be catastrophic under safety-critical scenarios \cite{li2017understanding}. In this paper, we restrict our scope to fail-continue errors from computing logic units (e.g., 1+1=3), assuming fail-stop errors are protected by checkpoint/restart and memory errors are protected by ECC. In what follows, we will use  {\it soft errors} to denote such fail-continue errors from computing logic units. 

Soft errors can be handled by dual modular redundancy (DMR). DMR approaches, typically assisted by compilers, duplicate computing instructions and insert check instructions into the original programs \cite{oh2002error, oh2002control, reis2005swift, yu2009esoftcheck, chen2016simd}. DMR is very general and can be applied to any application, but it introduces high overhead especially for computing-bound applications because it duplicates all computations. In order to reduce fault tolerance overhead, algorithm-based fault tolerance (ABFT) schemes have been developed for many applications in recent years. Huang and Abraham proposed the first ABFT scheme for matrix-matrix multiplication \cite{huang1984algorithm}. Sloan et al. proposed an algorithmic scheme to protect conjugate gradient algorithms of sparse linear systems \cite{sloan2012algorithmic}. Sao and Vuduc explore a self-stabilizing FT scheme for iterative methods \cite{sao2013self}. Di and Cappello proposed an adaptive impact-driven FT approach to correct errors for a series of real-world HPC applications \cite{di2016adaptive}. Chien at al. proposed the Global View Resilience system, a library which enables applications to add resilience efficiently \cite{chien2015versioned}. Many other FT schemes have been developed for widely-used algorithms such as sorting \cite{li2019ft}, fast Fourier transforms (FFT) \cite{liang2017correcting, antola1992fault, tao1993novel}, iterative solvers \cite{chen2013online, tao2016new, chen2014extending}, and convolutional neural networks \cite{zhao2020algorithm}. Recently, the interplay among resilience, power and performance is studied \cite{zamani2019greenmm, tan2014survey, tan2015investigating}, revealing the strong correlation among these key factors in HPC.

Although numerous efforts have been made to protect scientific applications from soft errors, most routines in the Basic Linear Algebra Subprograms (BLAS) library remain unprotected. The BLAS library is a core linear algebra library fundamental to a broad range of applications, including weather forecasting\cite{skamarock2008description}, deep learning \cite{tensorflow2015-whitepaper, NEURIPS2019_9015}, molecular dynamics simulation\cite{phillips2005scalable} and quantum computer simulation \cite{tang2020cutqc}. Because of this pervasive usage, academic institutions and hardware vendors provide a variety of BLAS libraries such as Intel MKL \cite{intel-alt}, AMD ACML, IBM ESSL, ATLAS \cite{whaley2001automated}, BLIS \cite{BLIS1}, and OpenBLAS \cite{wang2013augem} to pursue extreme performance on a variety of hardware platforms. BLAS routines are organized into three levels: Level-1 (vector/vector), Level-2 (matrix/vector), and Level-3 (matrix/matrix). \cite{whaley1998automatically}. Except for the general matrix-matrix multiplication (GEMM) routine, which has been extensively studied\cite{huang1984algorithm, wu2013line, gunnels2001fault, chen2008algorithm, smith2015toward}, minimal research has concentrated on protecting the rest of the BLAS routines.

For the general matrix-matrix multiplication routine, several fault tolerance schemes have been proposed to tolerate soft errors with low overhead \cite{huang1984algorithm, wu2013line, gunnels2001fault, smith2015toward}. The schemes in \cite{huang1984algorithm} and \cite{gunnels2001fault} are much more efficient than DMR. However, these two schemes are offline schemes which cannot correct errors in the middle of the computation in a timely manner. In \cite{wu2013line}, Wu et al. implemented a fault tolerant GEMM that corrects soft errors online. However, built on third-party BLAS libraries, this ABFT scheme becomes less efficient when using AVX-512-enabled processors because the gap between computation and memory transfer speed today becomes so large that the added memory-bound ABFT checksum computation is no longer negligible to the original computing-bound GEMM routine. In \cite{smith2015toward}, Smith et al. proposed a fused ABFT scheme for BLIS GEMM at assembly level to reduce the overhead for checksum calculation.  An in-memory checkpoint/rollback scheme is used to correct multiple simultaneous errors online. Although this scheme provides wider error coverage, it presents a moderate overhead ``in the range of 10\%”\cite{smith2015toward}.

In this paper, we develop FT-BLAS---the first BLAS implementation that not only corrects soft errors online, but also provides at least comparable performance to modern state-of-the-art BLAS libraries such as Intel MKL, OpenBLAS, and BLIS. FT-BLAS not only protects the general matrix-matrix multiplication routine GEMM, but also protects other Level-1, Level-2, and Level-3 routines. BLAS routines are widely-used in many applications from an extensive range of fields; therefore, improvements to the BLAS library will benefit not only a large number of people but also a broad cross-section of research areas. The main contributions of this paper include:

\begin{itemize}[leftmargin=*]
   \item We develop a brand-new implementation of BLAS using AVX-512 assembly instructions that achieves comparable or better performance than the latest versions of OpenBLAS, BLIS, and MKL on AVX-512-enabled processors such as Intel Skylake and Cascade Lake.
   \item We benchmark our hand-tuned BLAS implementation on an Intel Skylake processor and find that it is faster than the open-source librarues OpenBLAS and BLIS by 3.85\%-22.19\% for DSCAL, DNRM2, DGEMV, DTRSV, and DTRSM, and comparable ($\pm1.0\%$) for the remaining selected routines. Compared to closed-source Intel MKL, our implementation is faster by 3.33\%-8.06\% for DGEMM, DSYMM, DTRMM, DTRSM, and DTRSV, with comparable performance in the remaining benchmarks.
   \item We build FT-BLAS, the first fault-tolerant BLAS library, on our brand-new BLAS implementation by leveraging the hybrid features of BLAS: adopting a DMR strategy for memory-bound Level-1 and Level-2 BLAS routines and ABFT for computing-bound Level-3 BLAS routines. Our fault-tolerant mechanism is capable of not only detecting but also correcting soft errors online, during computation. Through a series of low-level optimizations, we manage to achieve a negligible (0.35\%-3.10\%) overhead.
   \item We evaluate the performance of FT-BLAS under error injection on both Skylake and Cascade Lake processors. Experimental results demonstrate FT-BLAS maintains a negligible performance overhead under hundreds of errors injected per minute while outperforming state-of-the-art BLAS implementations OpenBLAS, BLIS, and Intel MKL by up to 22.14\%, 21.70\% and 3.50\% , respectively---all of which cannot tolerate any errors.

\end{itemize}

 The rest of the paper is organized as follows: We introduce background and related works in Section II, and then detail how we achieve higher performance than the state-of-the-art BLAS libraries in Section III. Section IV and Section V present the design and optimization of our fault-tolerant schemes. Evaluation results are given in Section VI. We present our conclusions and future work in Section VII.

\section{Related Work and Background}
\subsection{Algorithm-Based Fault Tolerance}
 Algorithmic approaches to soft error protection for computing-intensive or iterative applications have achieved great success \cite{chen2008algorithm, chen2013online, wu2014ft, liang2017correcting, wu2016towards, smith2015toward, chen2016online, wu2016algorithm}, ever since the first algorithmic fault tolerance scheme  for matrix/matrix multiplication in 1984 \cite{huang1984algorithm}. The basic idea is that for a matrix-matrix multiplication $C=A\cdot B$, we first encode matrices into checksum forms. Denoting $e$=[$1,1,\dots,1$]$^T$, we have $A\xrightarrow[]{encode}A^c:=\begin{bmatrix}A \\ e^TA \end{bmatrix}$ and $B\xrightarrow[]{encode}B^r:=\begin{bmatrix}B & Be\end{bmatrix}$. With $A^c$ and $B^r$ encoded, we automatically have: $$ C^f = A^c\cdot B^r = \begin{bmatrix}C & Ce\\ e^TC & \end{bmatrix} = \begin{bmatrix}C & C^r\\ C^c & \end{bmatrix}$$ 
 The correctness of the multiplication can be verified by checking the matrix $C$ against $C^r$ and $C^c$. Any disagreements, that is, if the difference exceeds the round-off threshold, indicate errors occurred during the computation. The cost of checksum encoding and verification is $O(n^2)$, negligible compared to the $O(n^3)$ of matrix multiplication algorithms and thus ensures lightweight soft error detection for matrix multiplication. For any arbitrary matrix multiplication algorithm, correctness can be verified at the end of the computation (offline) via the checksum relationship. 
 
 The previous ABFT scheme can be extended to outer-product matrix-matrix multiplication and the checksum relationship can be maintained during the middle of computation:
 $$C^f = \sum_sA^c(:,s)\cdot B^r(s,:) = \sum_s\begin{bmatrix}C_s & C_se\\ e^TC_s & \end{bmatrix}$$
 where $s$ is the step size of the outer-product update on matrix $C$, and $C_s$ represents the result of each step of the outer-product multiplication $A^c(:,s)\cdot B^r(s,:)$. Noting this outer-product extension, Chen et al. proposed correcting errors for GEMM online with a double-checksum scheme \cite{chen2008algorithm}. The offline version of the double-checksum scheme can only correct a single error in a full execution, while the online version, which corrects a single error for \textit{each} step of the outer-product update, is able to handle multiple errors for the whole program. A checkpoint-rollback technique can also be added to overcome a many-error scenario. In \cite{smith2015toward}, once errors, regardless how many, are detected via the checksum relationship, the program restores from a recent checkpoint to correct the error. In this paper, we target a more light-weight error model and correct one error in each verification interval using online ABFT without checkpoint/rollback for the sake of performance.

\subsection{Duplication-Based Fault Tolerance}
Known as dual modular redundancy (DMR), duplication-based fault tolerance is rooted in compiler-assisted approaches and has been widely studied \cite{oh2002error, oh2002control, reis2005swift, yu2009esoftcheck, chen2016simd}. Classified by the Sphere of Replication (SoR), that is, the logical domain of redundant execution \cite{reinhardt2000transient}, previous duplication-based fault-tolerant work can be grouped into one of three cases:
\begin{itemize}[leftmargin=*]
\item Thread Level Duplication (TLD). This approach duplicates the entire processor and memory system: Everything is loaded twice, computed twice, and two copies are stored \cite{oh2002error, oh2002control}.
\item TLD with ECC assumption (TLD+ECC). In this approach, operands are loaded twice, but from the same memory address. All other instructions are still duplicated. \cite{reis2005swift}.
\item DMR only for computing errors. Only the computing instructions are duplicated and verified to prevent a faulty result from being written back to memory \cite{yu2009esoftcheck, chen2016simd}.
\end{itemize}
Different SoRs target different protection purposes and error models. TLD and TLD+ECC lead to the worst performance and memory overheads, but provide the best fault coverage without requiring any other fault-tolerance support such as checkpoint/restarting. Duplicating only the computing instructions shrinks the SoR to soft errors but almost halves the performance loss compared with TLD. We adopt the third SoR, duplication and verification of computing instructions only, in this work. 

Since compiler front ends never intrude into the assembly kernels of performance-oriented BLAS libraries, in the few cases that can be found in compiler literature relating to soft error resilience in BLAS routines \cite{chen2016simd}, the performance is \textsl{never} compared against OpenBLAS or Intel MKL, but only to  LAPACK \cite{laug}, a reference implementation of BLAS with much slower performance on modern processors. In this work, we manually insert FT instructions into self-implemented assembly computing kernels for Level-1 and Level-2 BLAS, and then hand-tune them for highest performance.  

\section{Optimizing Level-1, Level-2 and Level-3 BLAS routines}
Before adding FT capabilities to BLAS, we first create a brand new library that provides \textsl{comparable or better} performance to modern state-of-the-art BLAS libraries. We introduce the target instruction set of our work, as well as a sketch of the overall software organization. We then dive into our detailed optimization strategies for the assembly kernel to illustrate how we push our performance from the current state-of-the-art closer to the limits of hardware.

\subsection{Optimizing Level-1 BLAS}
Level-1 BLAS contains a collection of memory-bound vector/vector dense linear algebra operations, such as vector dot products and vector scaling.

\subsubsection{Opportunities to Optimize Level-1 BLAS}
Software strategies to optimize serial Level-1 BLAS vector routines are typically no more than exploiting data-level parallelism using vectorized instructions: processing multiple packed data via a single instruction, loop unrolling to benefit pipelining and exploit instruction-level parallelism, and inserting prefetching instructions. In contrast to computing-bound Level-3 BLAS routines, where performance can reach about 90\% of the theoretical limit, sequential memory-bound routines usually reach 60\%-80\% saturation because throughput is not high enough to hide memory latency. This fluctuating saturation range makes experimental determination of underperforming routines difficult. 
We therefore survey open-source BLAS library Level-1 routines source code with regard to three key optimization aspects: single-instruction multiple-data (SIMD) instruction set support, loop unrolling, and software prefetching. We include double-precision routines in Table \ref{tab:openblas-level-1} for analytical reference.

As seen in Table \ref{tab:openblas-level-1}, all Level-1 OpenBLAS routines have been implemented with support for loop unrolling. We also observe the interesting fact that software prefetching, an optimization strategy as powerful as increasing SIMD width for Level-1 routines, is only adopted in legacy implementations of x86 kernels in OpenBLAS. Based on the results of this optimization survey, we optimize two representative routines: we upgrade DNRM2 with AVX-512 support and enable prefetching for DSCAL. In the evaluation section, we show that the performance of our AVX-512-enabled DNRM2 with software prefetching surpasses OpenBLAS DNRM2 (SSE+prefetching) by 17.89\%, while our DSCAL with data prefetch enabled via \verb|prefetcht0| obtains a 3.85\% performance improvement over OpenBLAS DSCAL (AVX-512 with no prefetch).

\begin{table}[ht]
\centering
\begin{tabular}{|c|c|}
\hline
\textbf{AVX-512/AVX2}   & DDOT, DSCAL, DAXPY, DROT           \\ \hline
\textbf{AVX or earlier} & DNRM2, DCOPY, DROTM, IDAMAX, DSWAP \\ \hline
\textbf{Loop Unrolling} & all routines                       \\ \hline
\textbf{Prefetching}    & DNRM2, DCOPY, DROTM, IDAMAX, DSWAP \\ \hline
\end{tabular}
\caption{Survey of Selected OpenBLAS Level-1 Routines}
\label{tab:openblas-level-1}
\end{table}

\subsection{Optimizing Level-2 BLAS}

Level-2 BLAS performs various types of memory-bound matrix/vector operations. In contrast to Level-1 BLAS, which never re-uses data, register-level data re-use emerges in Level-2 BLAS. We choose the two most typical routines, DGEMV and DTRSV, as examples to explain the theoretical underpinnings of our Level-2 BLAS optimization strategies.

\subsubsection{Optimizing DGEMV}
DGEMV, double-precision matrix/vector multiplication, computes $ y=\alpha op(A)x + \beta y $, where $A$ is an $m\times n$ matrix and $op(A)$ can be $A$, $A^H$ or $A^T$. The cost of vector scaling $\beta y$ and $\alpha\cdot(Ax)$ is negligible compared with $A\cdot x$, therefore it suffices for us to consider $\beta = 1$, and $\alpha = 1$, and restrict our discussion to the case $y=Ax+y$, where $A$ is an $n\times n$ square matrix. The naive implementation can be summarized as $\sum^n_i(y_i=\sum^n_jA_{ij}x_j+y_i)$. Since DGEMV is a memory-bound application, the most efficient optimization strategy is to reduce unnecessary memory transfers. It is clear that the previous naive implementation requires $n^2$ loads for $A$, $x$ and $n^2$ loads + stores for $y$. No memory transfer operations can be eliminated on matrix $A$ because each element must be accessed at least one time. We must focus on register-level re-use for vectors $x$ and $y$ to optimize DGEMV. We notice that index variable $i$ in $A(i,j)$ is partially independent of the index $j$ of the j-loop, and we can unroll the i-loop $R_i$ times to exploit loading $x_j$ into registers for re-use. Now each load of $x_j$ is reused $R_i$ times within a single register, so the total load operations for $x$ improves from $n^2$ to $n^2$/$R_i$. In practice, $R_i$ is typically between 2-6, because accessing too many discontinuous memory addresses increases the likelihood of translation lookaside buffer (TLB) and row buffer thrashing. We adopt $R_i$=$4$ because the longest SIMD ALU instruction (\verb|VFMA|) latency in this loop is 4 cycles \cite{instructiontable}.

\begin{figure}[ht]
\centering
\scalebox{0.9}{\includegraphics[width=0.45\textwidth]{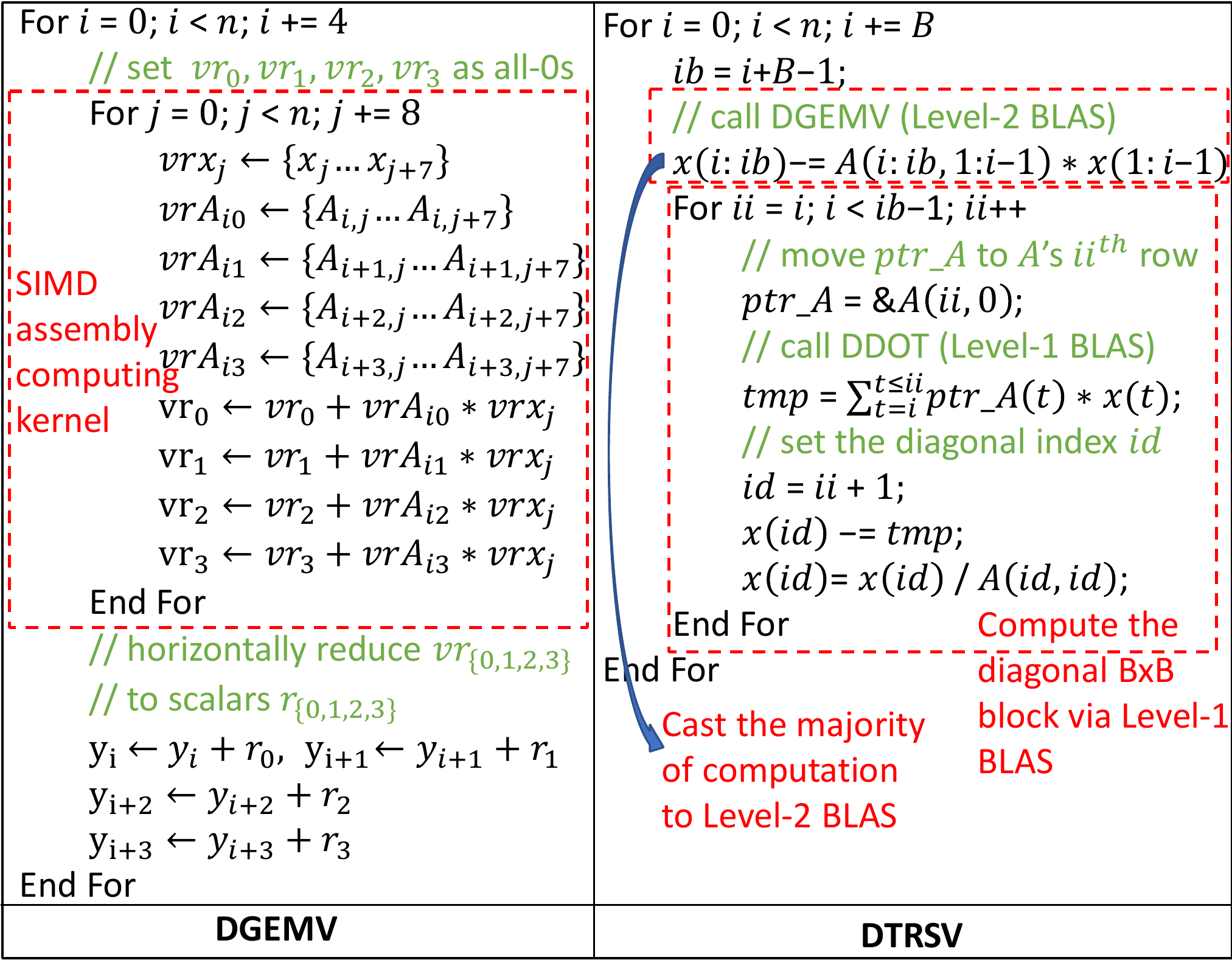}
}
\caption{Optimization schemes of DGEMV and DTRSV.}
\label{fig:loop-unroll}
\end{figure}

Unrolling the inner loop (j-loop) improves nothing in terms of load/store numbers, but will benefit a SIMD implementation (vectorization). Because both an AVX-512 SIMD register and a cache line of the Skylake microarchitecture accommodate 8 doubles, we unroll the j-loop 8 times. Before entering the j-loop, four SIMD registers $vr_{\{0,1,2,3\}}$ are initialized to zero. Within the innermost loop body, each $x$ element is still reused $R_i$ times (shown as 4 in Figure \ref{fig:loop-unroll}). We load 8 consecutive $x$ elements into a single SIMD AVX-512 register $vrx_j$, load the corresponding $A$ elements into SIMD registers $vrA_{i*}$, and conduct vectorized fused multiplication/addition operations to update $vr_*$. After exiting the j-loop, vectorized registers $vr_*$ holding temporary results are reduced horizontally to scalar registers, added onto the corresponding $y_i$, and stored back to memory. 
Some previous literature \cite{wang2013augem,whaley1998automatically} suggests blocking for cache level re-use of vector elements. However, this may break the continuous access of the matrix elements, which is the main workload of the DGEMV computation. Hence, we do not adopt a cache blocking strategy in our DGEMV implementation: experimental results validating our DGEMV obtain a 7.13\% performance improvement over OpenBLAS.

\subsubsection{Optimizing DTRSV} Double-precision triangular matrix/vector solver (DTRSV) solves $x=op(A)^{-1}x$, where A is an $n\times n$ matrix, $op(A)$  can be $A$, $A^H$ or $A^T$, and either the lower or upper triangular part of the matrix is used for computation due to symmetry. We restrict our discussion to $x=A^{-1}x$ using the lower triangular part of $A$. Since Level-2 BLAS routines are more computationally intensive than Level-1 BLAS routines, we introduce a paneling strategy for DTRSV to cast the majority of the computations --- $(n^2-nB)/2$ elements --- to the more computationally-intensive Level-2 BLAS routine DGEMV. The minor $B\times B$ diagonal section is handled with the less computationally-intensive Level-1 BLAS routine DDOT. Given that DGEMV is more efficient, adopting a smaller block size $B$ is preferable since it allows more computations to be handled by DGEMV. Considering the practical implementation of DGEMV, where we unroll the j-loop 4 times for register re-use (shown in Figure \ref{fig:loop-unroll}), the minimal, and also the optimal, block size $B$ should then be 4. In fact, OpenBLAS adopts block size $B$=$64$ for DTRSV \cite{openblasdtrsv}, resulting in more computations handled by the less efficient diagonal routine; this is the major reason our performance supersedes that of OpenBLAS by 11.17\%.

\subsection{Optimizing Level-3 BLAS}
\subsubsection{Overview of Level-3 BLAS}
Level-3 BLAS routines are matrix/matrix operations, such as dense matrix/matrix multiplication and triangular matrix solvers, where extreme cache and register level data re-use can be exploited to push the performance to the peak computation capability. We choose two representative routines, DGEMM and DTRSM to illustrate our implementation and optimization strategies for Level-3 BLAS.

\subsubsection{Implementation of DGEMM} We adopt packing and cache blocking frames similar to OpenBLAS and BLIS. The outermost three layers of the \verb|for| loop are partitioned to allow submatrices of $A$ and $B$ to reside in specific cache layers. The step sizes of these three \verb|for| loops, $M_C$, $N_C$, and $K_C$, define the size and shape of the macro kernel, which are determined by the size of each layer of the cache. A macro kernel updates an $M_C\times N_C$ submatrix of C by iterating over $A$ $(M_R\times K_C)$ multiplying $B$ $(K_C\times N_R)$ in micro kernels. Since our implementation contains no major update on the latest version of OpenBLAS other than selecting different micro kernel parameters $M_R$ and $N_R$, nor on the performance ($<\pm0.5\%$), we do not present a detailed discussion of the DGEMM implementation here but instead refer readers to \cite{BLIS1} for more details.

\subsubsection{Optimizing DTRSM} DTRSM, double-precision triangular matrix/matrix solver, solves $B=\alpha\cdot op(A)^{-1}B$ or $B=\alpha B\cdot op(A)^{-1}$, where $\alpha$ is a double-precision scalar, $A$ is an $n\times n$ matrix, $op(A)$  can be $A$, $A^H$, or $A^T$, and either the lower or upper triangular part of the matrix is used for computation due to symmetry. We restrict our discussion to $B=A^{-1}B$ in the presentation of our optimization strategy. We adopt the same cache blocking and packing scheme as DGEMM, but with the packing routine for $A$ and the macro kernel slightly modified. For DTRSM, the packing routine for matrix $A$ not only packs the matrix panels into continuous memory buffers to reduce TLB misses, but also stores the reciprocal of the diagonal elements during the packing to avoid expensive division operations in the performance-sensitive computing kernels. When the $A_{block}$ to feed into the macro kernel is on the diagonal, \verb|macro_kernel_trsm| is called to solve $B_{block}$ := $\Tilde{A}^{-1}\cdot \Tilde{B}$, where $\Tilde{A}$ and $\Tilde{B}$ are packed matrices. Otherwise, the corresponding $B_{block}$ is updated by calculating $B_{block}$ -= $\Tilde{A}\cdot \Tilde{B}$, using the highly-optimized GEMM macro kernel. We see that the performance of the overall routine is affected by both macro kernels, and to ensure overall high performance, we must ensure the TRSM kernel is near-optimal as well.

Inside \verb|macro_kernel_trsm|, the $B_{block}$ is calculated by updating a small $M_R \times N_R$ $B_{sub}$ block each time. The $B_{sub}$ block is calculated by $B_{sub}$ -=  $\Tilde{A_{curr}}\cdot \Tilde{B_{block}}$ until $A_{curr}$ reaches the diagonal block. Temporary computing results are held in registers instead of being saved to memory during computation. When $A_{curr}$ is on the diagonal, we solve $B_{sub}$ :=  $\Tilde{A_{curr}}^{-1}\cdot \Tilde{B_{block}}$ using an AVX-512-enabled assembly kernel. It should be noted that the packed buffer $\Tilde{B}$ needs to be updated during the solve because DTRSM is an in-place update and the corresponding elements of the buffer should be updated during computation. Our highly-optimized TRSM macro kernel grants us 22.19\% overall performance gain on DTRSM over OpenBLAS, where the TRSM macro kernel is an under-optimized prototype.

\begin{figure}[ht]
\vspace{-1mm}
\centering
\includegraphics[width=0.48\textwidth]{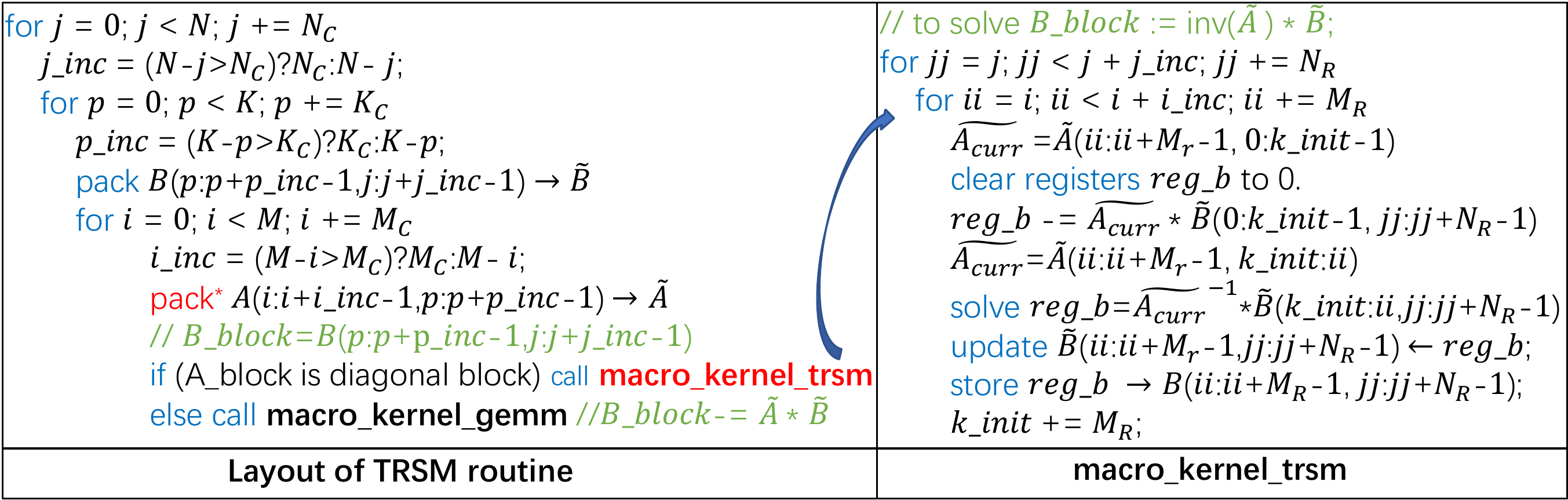}
\caption{DTRSM optimization layout.}
\label{fig:dtrsm-layout}
\vspace{-4mm}
\end{figure}


\section{Optimizing Fault Tolerant Level-1 and Level-2 BLAS}
\label{section:FT}
We first outline our assembly code syntax and duplication scheme. We then show our step-wise assembly optimization of DMR to decrease fault tolerance overhead from 50.8\% in the scalar version to our 0.35\% overhead. After the optimization, the performance of both our FT and non-FT versions surpasses both current state-of-the-art BLAS implementations.

\subsection{Assembly Syntax and Duplication Scheme}
 In this paper, all assembly examples follow AT\&T syntax; that is, the destination register is in the right-most position. We adopt the most common duplication scheme, DMR \cite{reis2005swift, yu2009esoftcheck, oh2002error}. Our chosen sphere of reduction dictates that we duplicate and verify computing instructions instead of memory instructions. More specifically, in our case, most ALU operations are floating point operations. Integer addition/subtraction are used to check whether the loop terminates. We only use two integer registers (\%0, \%1) throughout our assembly kernels.

\subsection{Scalar DMR versus Vectorized DMR}

We use DSCAL, one of the most important routines in Level-1 BLAS, to show how even though DMR is labeled ``slow", it can actually be ``fast". DSCAL computes $x:=\alpha\cdot x$, where $x$ is a vector containing $n$ DPs. DP represents a double-precision data type, so $\alpha$ is also a DP.
\subsubsection{Scalar scheme}

The scalar implementation of DSCAL performs a load (\verb|movsd|), multiplication (\verb|mulsd|), and then a store (\verb|movsd|) operation on scalar elements. The scalar $\alpha$ is invariant within the loop body, so we load it before entering the loop. The array index (stored in register \%0) to access array elements is incremented by \$1 before starting the next iteration. Meanwhile, register \%1 (initialized by the array length $n$) is decremented by one to test whether the loop terminates. Once register \%1 reaches zero, the EFLAG ZF is set to 1, branch instruction \verb|jnz| will not be taken, and the loop terminates. Because scalar multiplication \verb|mulsd| only supports two-operand syntax---that is, \verb|mulsd, src, dest| multiplies values from two operands and stores the result in the \verb|dest| register---the value in the \verb|dest| register will be overwritten when the computation finishes. Therefore, we should back up a copy of the loaded value of $x[i]$ into an unused register for use in our duplication to avoid an extra load from memory. After both the original and duplicated computations finish, we check for correctness and set the EFLAGs via \verb|ucomsid|. If two computing results (\verb|xmm1| and \verb|xmm2|) are different, the EFLAG is set as ZF=1 and the branch \verb|jne ERROR_HANDLER| will redirect the control flow to activate a resolving procedure, a self-implemented error handling assembly code. When the correctness of computing is confirmed or an erroneous result is recovered by the error handler, the result $\alpha\cdot x[i]$ is stored into memory.
\subsubsection{AVX-512 vectorized scheme}

Our AVX-512 vectorized duplication scheme differs from the scalar version in two ways. First, vectorized multiplication supports a three operand syntax, so source operand registers are still live after computing and an in-register backup is no longer needed. Second, comparison between SIMD registers cannot set EFLAGs directly. Therefore, we set EFLAGs indirectly: The comparison result is first stored in an opmask register \verb|k0|, and then \verb|k0| is tested against another pre-initialized opmask register \verb|k1| to set EFLAGs. If two 512-bit SIMD registers with 8 packed DPs are confirmed equal, opmask register \verb|k0|, updated by \verb|vpcmpegd|, will be eight consecutive `1's corresponding to the eight DPs in the comparison. If one (or more) DP(s) from two source operands in comparison are different, the corresponding bit(s) of the opmask register is set to \verb|0|, indicating the erroneous position. We test the comparison result opmask, \verb|k0|, with another opmask, \verb|k1|,  pre-initialized to \verb|0000|\verb|0000| via \verb|kortestw|. EFLAG is set to CF=0 first, and updated to CF=1 only if the results of OR-ing both source registers (\verb|k0, k1|) are all `\verb|1|'s. Any detected errors will leave CF=0, and the control flow is branched to the error handler by \verb|jnc|.
\subsubsection{Performance gain due to vectorization}
Our vectorized FT enlarges the verification interval compared to the scalar implementation: The scalar scheme gives a computing/comparison+branch ratio of 1:1, while the vectorized scheme expands this ration to 8:1, which significantly ameliorates the data hazards introduced by duplication and verification. Experimental results confirm that vectorization improves the overhead from 50.8\% in the scalar scheme to 5.2\% in the vectorized version.
\subsection{Adding More Standard Optimizations}
The peak single-core performance of an Intel processor that supports AVX-512 instructions is 30-120 GFLOPS, while the performance of DSCAL is less than 2 GFLOPS. Since CPU utilization is severely bounded by memory throughput, the inserted FT instructions, which do not introduce extra memory queries, should ideally bring a near-zero overhead if computations and memory transfers are perfectly overlapped. This underutilization of CPU performance motivates us to explore optimization strategies to further bring the current 5\% overhead to a negligible level.
\subsubsection{\textbf{Step 1}: Loop unrolling}
Loop unrolling is a basic optimization strategy for loop-based programs. However, it can only reduce a few branch and add/sub integer instructions in practice because CPUs automatically predict branches and unroll loops via speculative execution. Possible data hazards caused by speculative execution can be ameliorated by out-of-order execution mechanisms in hardware. Experiments show that the performance of both our FT and non-FT versions only slightly improves after unrolling the loop 4 times: The overhead decreases from 5.2\% to 4.9\%.
\subsubsection{\textbf{Step 2}: Adding comparison reduction}

Inspired by the previous ten-fold improvement on overhead due to the enlargement of the verification interval, this optimization is naturally focused on the reduction of branch instructions for comparison and diverging to the error handler by leveraging features of the AVX-512 instruction set. Intermediate comparison results are stored in opmask registers and a correct comparison result is stored as ``\verb|11111111|" in an opmask register, so we can propagate the comparison results via \verb|kandw k1, k2, k3|, AND-ing the two intermediate comparison results (\verb|k1,k2|), and storing into the third opmask register \verb|k3|. The AND operation ensures that any detected incorrectness marked by ``0" in source opmask registers will pollute bit(s) in the destination register during \textsl{reduction} and will be kept. 
Instead of inserting a branch to the error handler for each comparison, only one branch instruction is needed for every 4 comparisons in a loop iteration. This enlargement of the verification interval further decreases the overhead from 4.9\% to 2.7\%.

\subsection{Optimizations Underrepresented in Main Libraries}

We have still not reached optimality at this time. We review possible performance concerns left from the previous step:
\begin{itemize}[leftmargin=*]
    \item Data hazards. A read-after-write hazard is a true data dependency, and severely impacts this version of the code.
    \item Structural hazards. Four consecutive store instructions all demand specific AVX-512 units, but there are only two in SkylakeX processors; the instructions stall until hardware becomes available.
\end{itemize}
 Although out-of-order execution performed by a CPU can avoid unnecessary stalls in the pipeline stage, it consumes hardware resources and those resources are not unlimited. Therefore, we optimize instruction scheduling manually, assuming no hardware optimizations.
\subsubsection{Heuristic software pipelining}
We perform software pipelining to reschedule the instructions \textsl{across} the boundary of basic blocks in order to reduce structural and data hazards. Unfortunately, finding an optimal software pipelining scheme is NP-complete \cite{hsu1986highly}. To simplify the issue, we design the software pipelining heuristically by not considering the actual latency of each type of instruction. To scale eight consecutive elements that can be packed and processed in a AVX-512 SIMD register, we should first load them from memory (L), multiply with the scalar (M1), duplicate multiplication for verification (M2), compare between the original and duplicated results (C), and store back to memory (S) if correct. Stacking these five stages within the loop body causes a severe dependency chain because they all work on the same data stream. To deal with this issue, we first write down the required five stages for a single iteration (L, M1, M2, C, S) vertically and issue horizontally with a one-cycle latency for two adjacent instruction streams.
 
\begin{figure}[ht]
\centering
\includegraphics[width=0.46\textwidth]{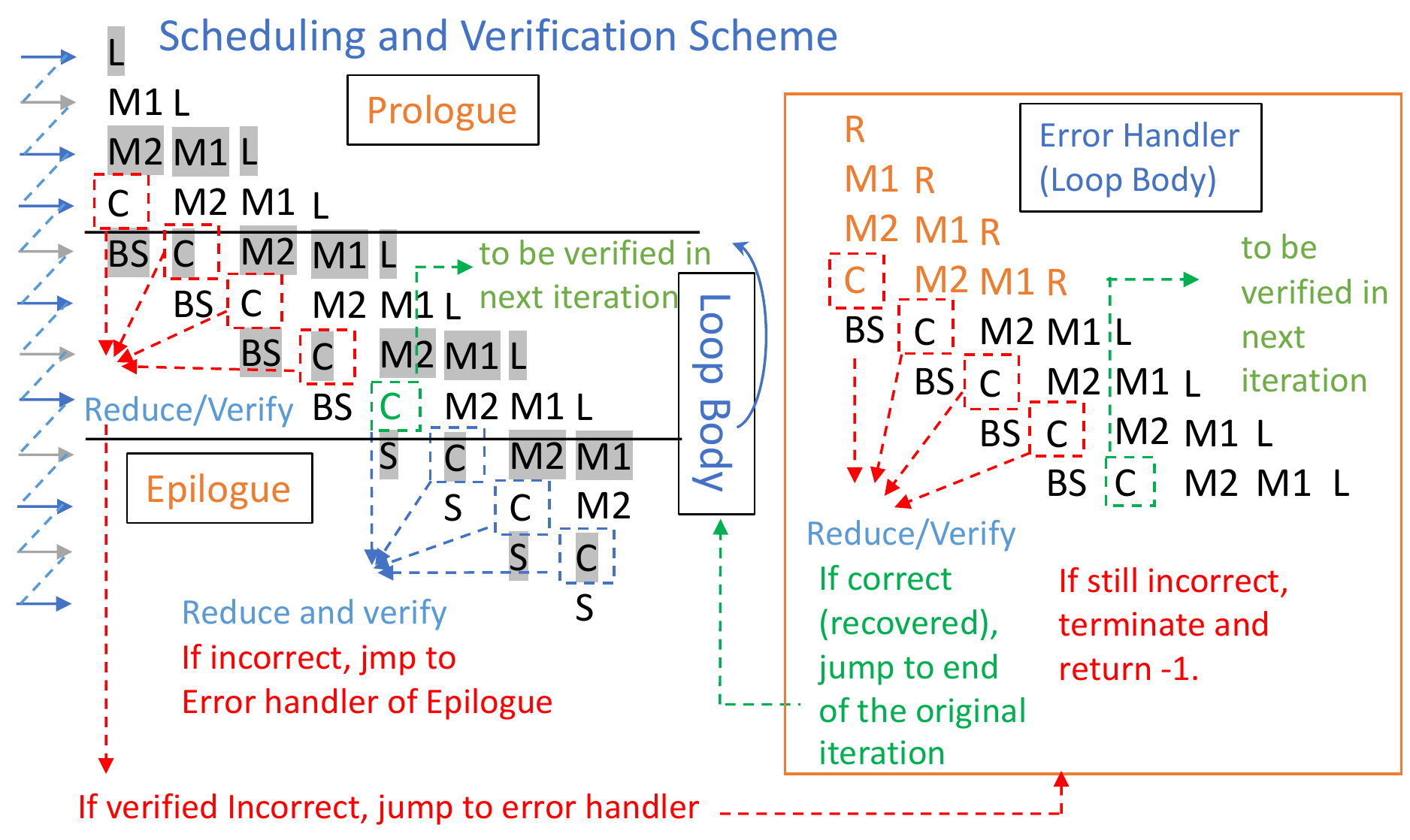}
\caption{Software pipelining design. \small{{\it \textmd{Each letter represents a vectorized instruction. L: Load; M1: Mul; M2: Duplicated Mul; C: Vectorized Comparison; S: Store; BS: Checkpoint original value before scaling into an unused register, then Store the computing result back to memory; R: Restore from a checkpoint register.}}}}
\end{figure}
 
 \subsubsection{Verification reduction and in-register check-pointing}

 Since the loop is still unrolled four times, comparison results can be reduced via \verb|kandw| between opmask registers. The next loop iteration will start to execute only if the loop does not terminate and the correctness of the current iteration is verified. With cross-boundary scheduling, we compute for iterations 2, 3, 4, 5 but verify iterations 1, 2, 3, 4. The comparison result of the fifth iteration is only stored, and then verified in the next iteration or in the epilogue. Because the memory is updated before computing results are verified, we checkpoint original elements loaded from memory in an unused register. This operation coalesces the ``in-register checkpoint" (B) followed by a store (S), and is denoted by BS when designing our software pipelining. Once an error is detected in the loop body and the recovery procedure is activated, the error handler restarts the computation from a couple of prologue-like instructions where the load is substituted with recovery from the backup registers. The corruption is recovered by a third calculation with duplication so the results must be verified again. If the disagreement still exists, the program is terminated and signals that it is unable to recover. If recovered computing results reach consensus, the control flow returns back to the end of corrupted loop iteration and continues as normal.
\subsubsection{Effectiveness of scheduling}

Experimental benchmarks report the latencies of  \verb|vmulpd|, \verb|vcmpeqpd|, and \verb|vmovpd| (both store and load) are 4, 3, and 3 cycles (under a cache hit), respectively \cite{instructiontable}. After scheduling, operands are consumed after 3 instructions; before our scheduling these operands are consumed immediately by the following instruction. For structural hazards, according to the Intel official development manual \cite{intel2019intel}, two adjacent vectorized multiplications (M2, M1) can be executed by Port 0 and Port 1, and Port 5 accommodates the following comparison (C) simultaneously. Therefore, three consecutive ALU operations $C, M2, M1$ within the loop body produce no structural hazard concerns. Additionally, Skylake processors can execute two memory operations at the same time so the structural hazard concerns on load and store are also eliminated. Therefore, we confirm that our heuristic scheduling strategy on DSCAL effectively ameliorates the hazards introduced by fault tolerance. We optimize the non-FT version using the same method and compare with our FT version. Experimental result demonstrates that software pipelining improves FT overhead from 2.7\% to 0.67\%.

\subsubsection{Adding software prefetching}

Prefetching data into the cache before it is needed can lower cache miss rates and hide memory latency. Dense linear algebra operations demonstrate high regularity on their memory access patterns, enabling performance improvement via accurate cache prefetching. We can send a prefetch signal \textsl{before} data is needed by a \textsl{proper prefetch distance}. When the data is actually needed, it has been prefetched into cache instead of waiting the approximately 100 ns required to load it from DRAM. Accurate prefetching distance is important. If data is prefetched too early or too late, the cache is polluted and performance can degenerate. Here we select the prefetch distance to be 1024 bits: We prefetch 128 elements in advance into the L1 cache using the instruction \verb|prefetcht0|. Instead of prefetching for all load operations, we only prefetch half of them in the loop body to avoid conflicts with hardware prefetching. Prefetching improves the performance of both our non-FT and FT versions by \textasciitilde$4\%$, and the overhead further decreases from 0.67\% to 0.36\%.

\section{Optimizing Fault Tolerant Level-3 BLAS}
Since Level-3 BLAS routines are computing-bound routines, adopting the same DMR strategy as Level-1 and Level-2 BLAS, which doubles the computing instructions, will consequently double the performance overhead. Considering the limited registers in a single core, DMR will also increase the register pressure in the computing kernels, which will further hinder the performance. Therefore, we adopt the classic checksum-based ABFT scheme for our fault-tolerant functionality, introducing $O(n^2)$ computational overhead over the original $O(n^3)$ computation.

\subsection{First trial: building online ABFT on a third-party library}
\begin{figure}[ht]
\centering
\includegraphics[width=0.48\textwidth]{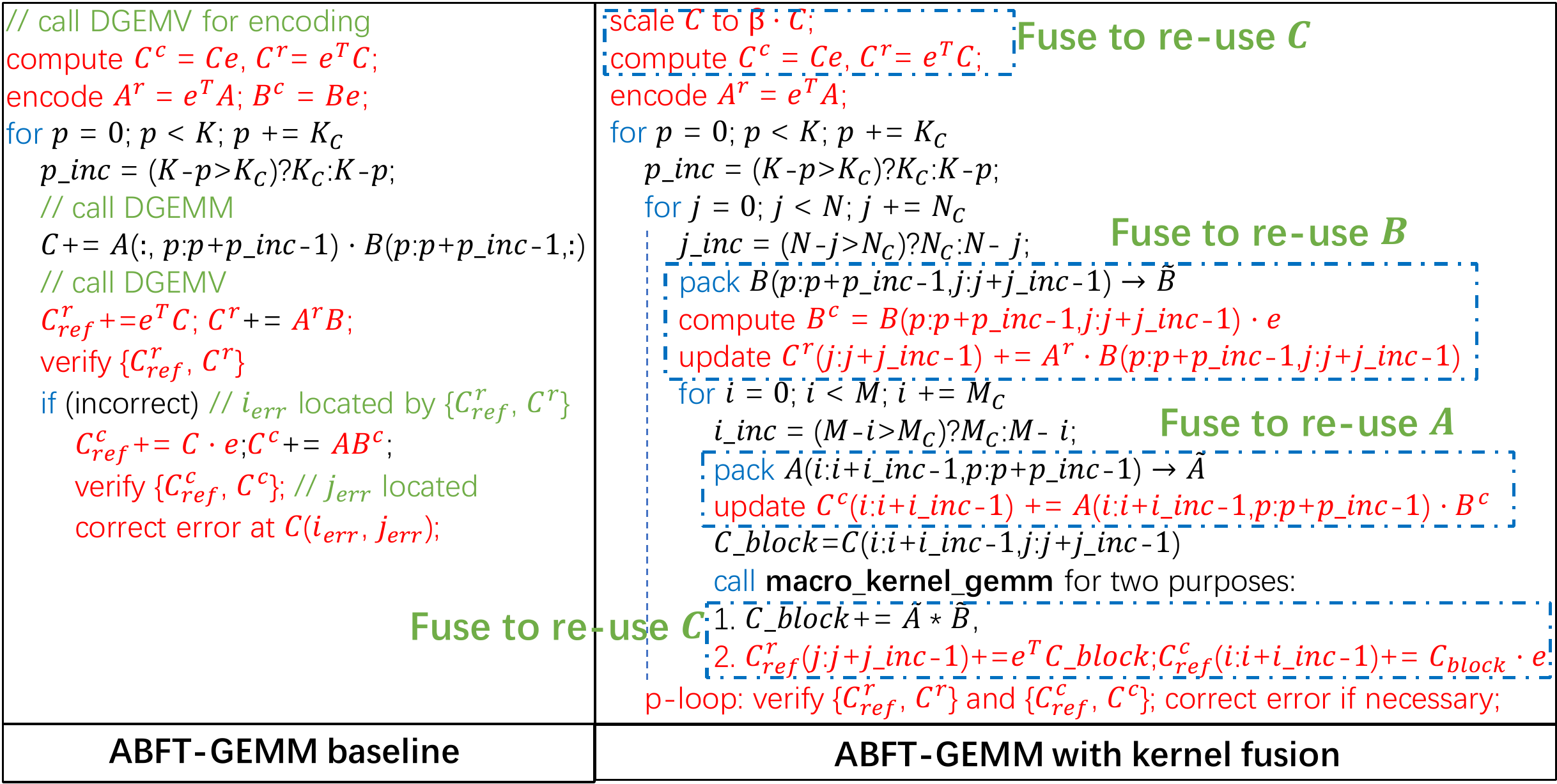}
\caption{outer-product online ABFT DGEMM optimization layout. \small{{\it \textmd{The ABFT-related operations are marked in red.}}}}
\label{fig:dgemm-layout}
\end{figure}

Building ABFT on a third-party library is not a new topic \cite{wu2013line}. As shown in the left side of Figure \ref{fig:dgemm-layout}, we first encode checksums for matrices $A$, $B$, and $C$ before starting matrix multiplication. The checksums $C^c$ and $C^r$ are updated asynchronously with the rank-k outer-product update of matrix $C$ with a step size $k$=$K_c$. In every completed rank-k update, we verify the checksum relationship by first computing the reference row checksum $C^r_{ref}$ according the current matrix $C$ and comparing it against $C^f$. If an error is detected, we continue to compute the reference column checksum $C^c_{ref}$ and compare against $C^c$ to locate the erroneous row index $i_{err}$ of $C$. If there is no error detected when comparing the row checksum vectors, we do not need to verify the column checksum vectors. 

The total cost of the ABFT overhead consists of the initial checksum encoding, online checksum updating, and reference checksum computing--all of which are matrix-vector multiplications (DGEMV). $T_{enc}$ includes the costs of encoding for four checksums ($C^c$,$C^r$,$A^r$,$B^c$). $T_{update}$ includes the costs of updating on two checksums ($C^c$,$C^r$). Denoting the time of an $n \times n$ DGEMV as $t_{mv}$, the total cost of ABFT $T_{ovhd}$ is:
$$T_{ovhd}=T_{enc}+T_{update}+\frac{K}{Kc}\cdot (T_{C^r_{ref}}+T_{C^c_{ref}}) = (6+\frac{2K}{Kc})t_{mv}$$

We further denote the performance of DGEMV and DGEMM as $P_{mv}$ and $P_{mm}$, both in the unit of GFLOPS. Then the total execution times of $n\times n$ DGEMM and DGEMV are $T_{GEMM}$=$2e^{-9}n^3/P_m$ and $t_{mv}$=$2e^{-9}n^2/P_{mv}$. Therefore, we have:
$$\frac{T_{ovhd}}{T_{GEMM}}=\frac{(6+\frac{2K}{Kc})t_{mv}}{2e^{-9}n^3/P_{mm}}=\frac{(6+\frac{2K}{Kc})P_{mm}}{n\cdot P_{mv}}$$
As shown in the above derivation, the real influence of ABFT is not simply $O(1/n)$ computationally negligible to the baseline, but dependent on the relative performance between the memory-speed-determined $P_{mv}$ and the computing-capability-determined $P_{mm}$ as well. On non-AVX-512-enabled CPUs, $P_{mm}/P_{mv}$ ranges from 5 to 20, while on AVX-512-enabled CPUs, this ratio can be as large as 35, exaggerating the overhead up to 7-fold over old processors. The ABFT overhead reported for an older CPU \cite{wu2013line} is around 2\%, while the overhead on an AVX-512-enabled processor, measured by our benchmark in Section VI, is 15.27\% --- much larger than on old processors.


\subsection{Reducing the memory footprint: fusing ABFT into DGEMM}
As discussed in the previous section, the huge gap between memory transfer and floating-point computation is the reason why the $O(n^2)$ checksum-related operations can no longer be amortized by $O(n^3)$ GEMM. We therefore design a fused ABFT scheme to minimize the memory footprint of checksum operations. To be more specific, the encoding of $C^c$ and $C^r$ is fused with the matrix scaling routine $C$=$\beta C$. When we load $B$ to pack it to the continuous memory buffer $\Tilde{B}$, checksum $B^c$ is computed and checksum $C^r$ is computed simultaneously by reusing $B$. In this fused packing routine, each $B$ element is reused three times for each load. Similarly, each element of $A$ loaded for packing is reused to update the column checksum $C^c$. In the macro kernel, which computes $C_{block}$+=$\Tilde{A}\cdot\Tilde{B}$, we reuse the computed $C$ elements at register level to update the reference checksums $C^r_{ref}$ and $C^c_{ref}$ in order to verify the correctness of the computation. By fusing the ABFT memory footprint into DGEMM, the FT overhead becomes purely computational, decreasing from about 15\% to 2.94\%.


\section{Experimental Evaluation}
\label{section:eva}

To validate the effectiveness of our optimizations, we compare the performance of FT-BLAS with three state-of-the-art BLAS libraries, Intel oneMKL (\verb|2020.2|, abbreviated as MKL in this Section), OpenBLAS (\verb|0.3.13|), and BLIS (\verb|0.8.0|), on a machine with an Intel Gold 5122 Skylake processor at 3.60 GHz, equipped with 96 GB DDR4-2666 RAM. We also compare the performance of FT-BLAS under error injection with references on an Intel Xeon W-2255 Cascade processor. This Cascade Lake machine has a 3.70 GHz base frequency and 32 GB DDR4-2933 RAM. Hardware prefetchers on both machines are enabled according to the Intel BIOS default \cite{prefetcher}. We repeat each measurement twenty times and then report the average performance. For Level-1 BLAS routines, the performance is averaged from array lengths ranging from $5\times10^6$ to $7\times10^6$. For Level-2 and Level-3 BLAS routines, the performance is averaged for matrices ranging from $2048^2$ to $10240^2$. We compile the code with \verb|icc 19.0| and the optimization flag \verb|-O3|.
\subsection{Performance of FT-BLAS without FT Capability}
We provide a brand-new BLAS implementation, comparable or faster than the modern state-of-the-art, before embedding FT capability. We abbreviate this BLAS implementation \textit{FT-BLAS: Ori} in the figures.
\subsubsection{Optimizing Level-1 BLAS}
For memory-bound Level-1 BLAS, the optimization strategies employed are: 1) exploiting data-level parallelism using the latest SIMD instructions, 2) assisting pipelining by unrolling the loop, and 3) prefetching. As seen in Table \ref{tab:openblas-level-1}, OpenBLAS has under-optimized routines, such as DSCAL and DNRM2, with respect to prefetching and AVX-512 support. We add prefetching for DSCAL, obtaining 3.85\% and 5.61\% speed-up over OpenBLAS and BLIS. DNRM2 is only supported with SSE2 by OpenBLAS, so our AVX-512 implementation provides a 17.89\% improvement over OpenBLAS, while reaching 2.25-fold speedup on BLIS. Our implementations for both routines reach comparable performance to closed-source MKL, as seen in Figure \ref{fig:other-level-1}.

\subsubsection{Optimizing Level-2 BLAS}

\begin{figure}[ht] \centering
\subfigure[DSCAL]
{
\includegraphics[width=0.21\textwidth]{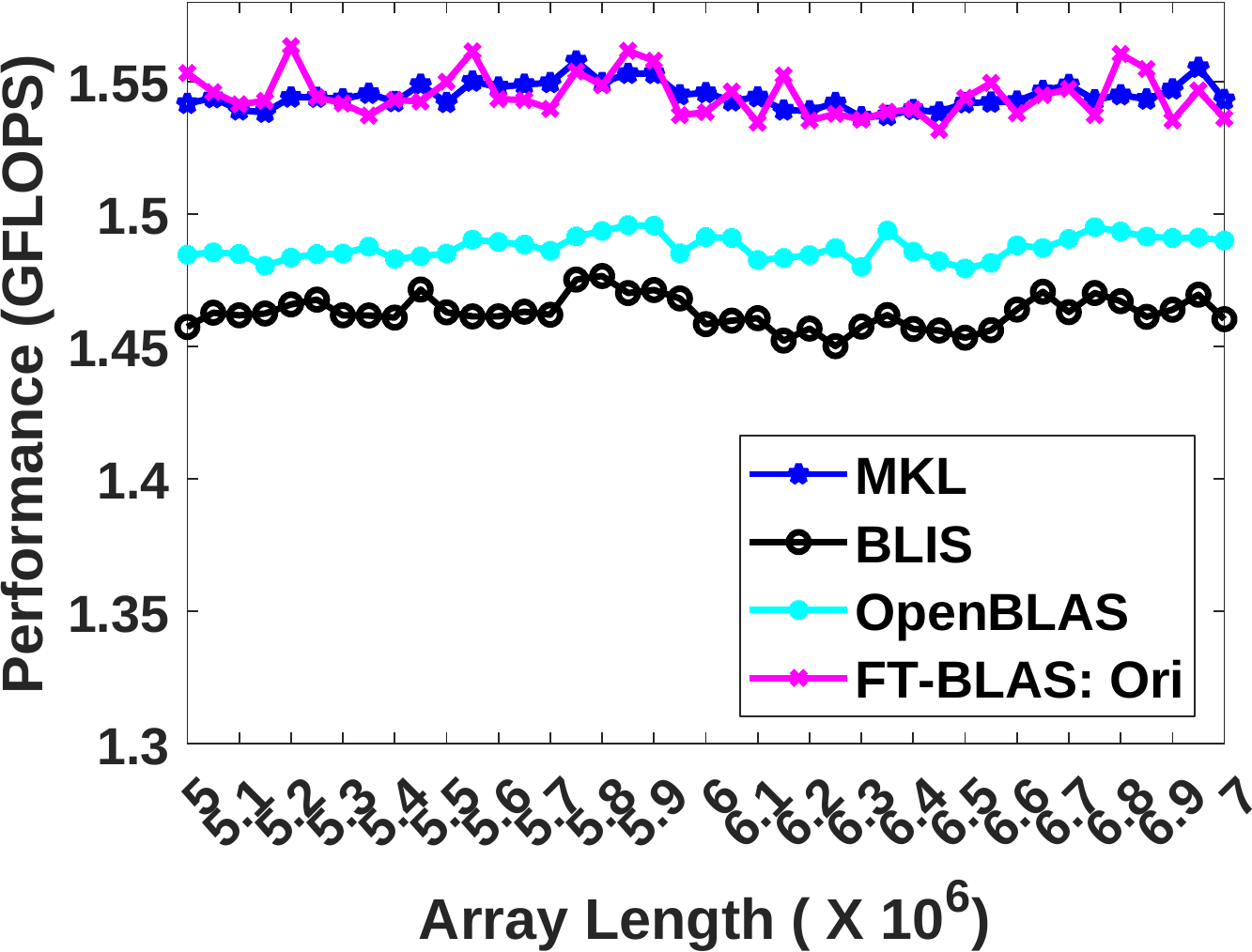}
}
\hspace{-3mm}
\vspace{-3mm}
\subfigure[DNRM2]
{
\includegraphics[width=0.21\textwidth]{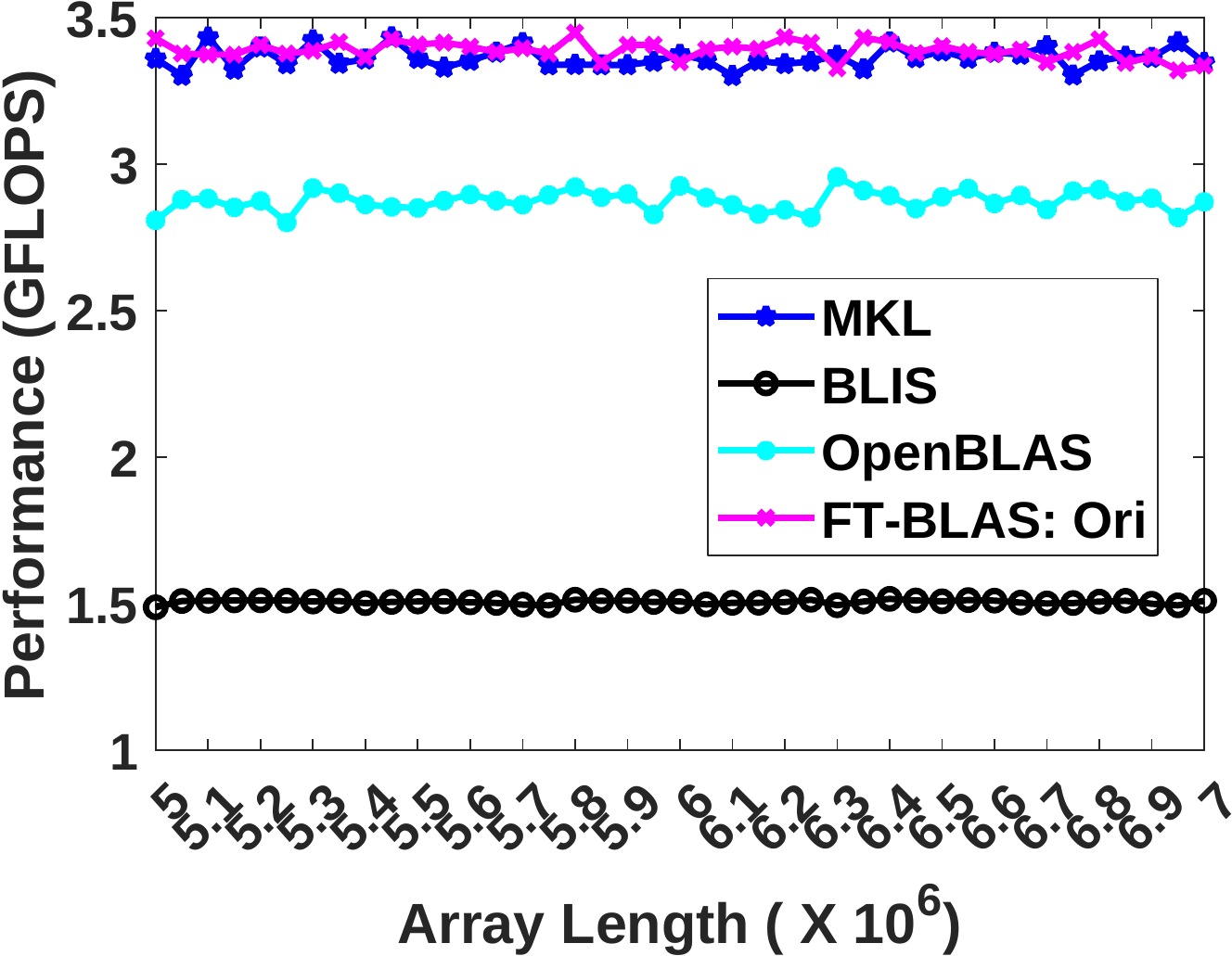}
}
\hspace{-3mm}
\vspace{-3mm}
\subfigure[DGEMV]
{
\includegraphics[width=0.21\textwidth]{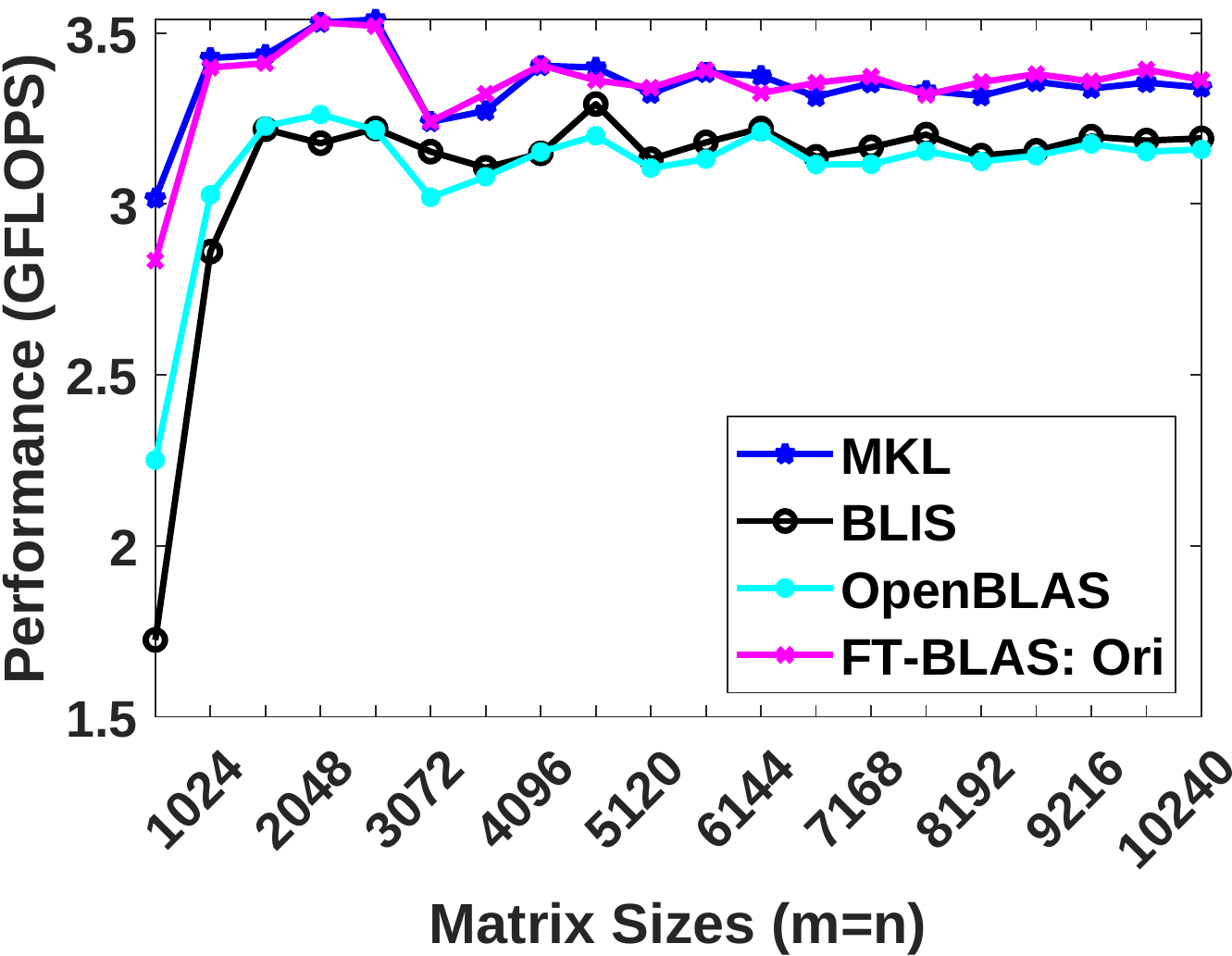}
}
\hspace{-3mm}
\vspace{-1mm}
\subfigure[DTRSV]
{
\includegraphics[width=0.21\textwidth]{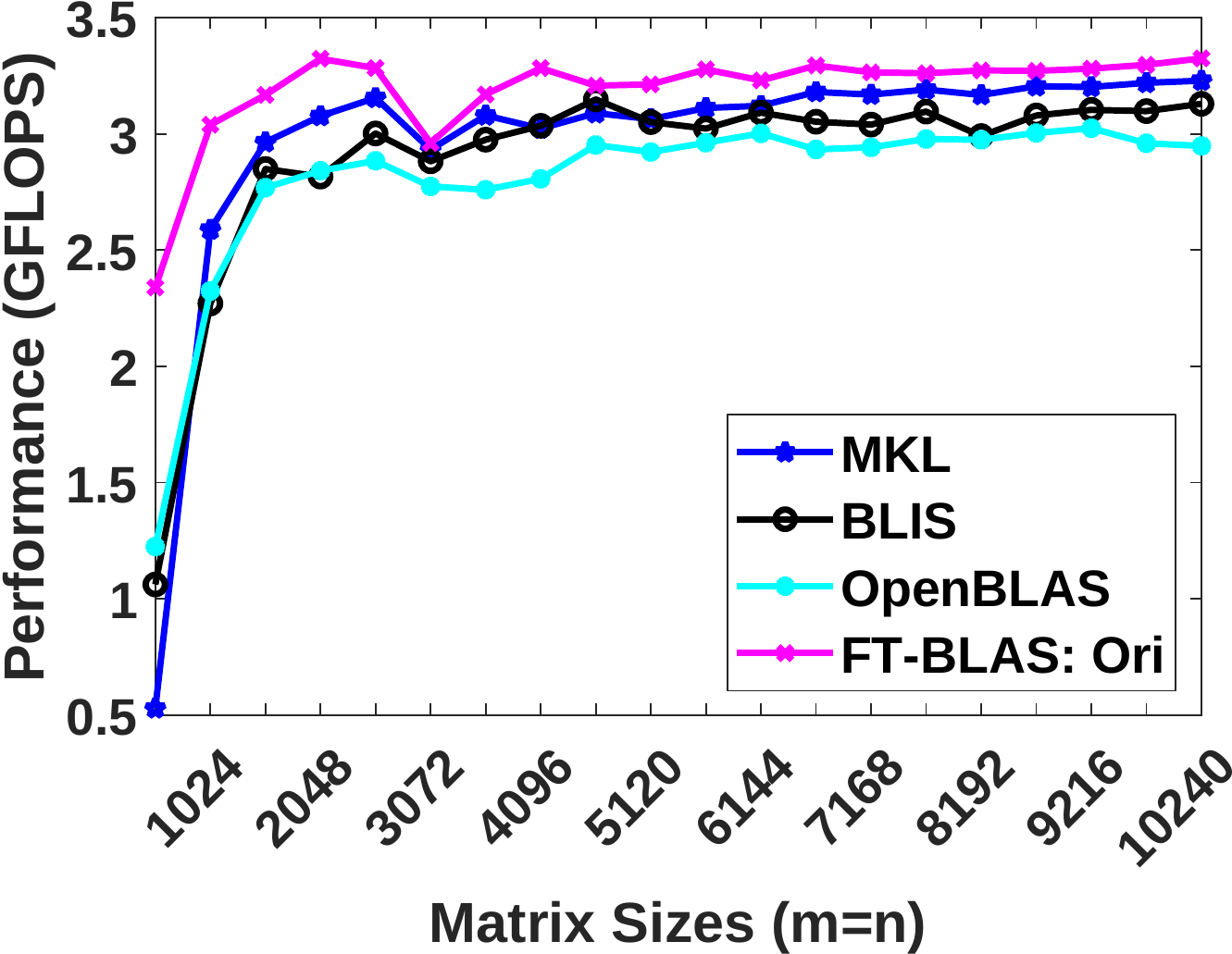}
}
\caption{Comparisons of selected Level-1/2 BLAS routines.}
\label{fig:other-level-1}
\vspace{-4mm}
\end{figure}

Register-level data re-use enters the picture in Level-2 BLAS routine optimization. Following the optimization schemes described in Section II, we see in Figure \ref{fig:other-level-1} that our DGEMV obtains a 7.13\% speed-up over OpenBLAS by discarding cache blocking on matrix $A$ over concerns about the potential harm of discontinuous memory accesses regarding TLB thrashing and the corresponding performance of hardware prefetchers. Because BLIS adopts the same strategy as OpenBLAS on DGEMV, our DGEMV is 6.16\% faster than BLIS, while achieving nearly indistinguishable performance with MKL. For DTRSV, our strategy of minimizing the blocking parameter to cast the maximized computations to the more efficient Level-2 BLAS DGEMV grants us higher performance than all baselines, surpassing MKL, OpenBLAS, and BLIS by 3.76\%, 11.17\%, and 6.98\%, respectively.

\begin{figure}[ht] \centering
\subfigure[DGEMM]
{
\includegraphics[width=0.21\textwidth]{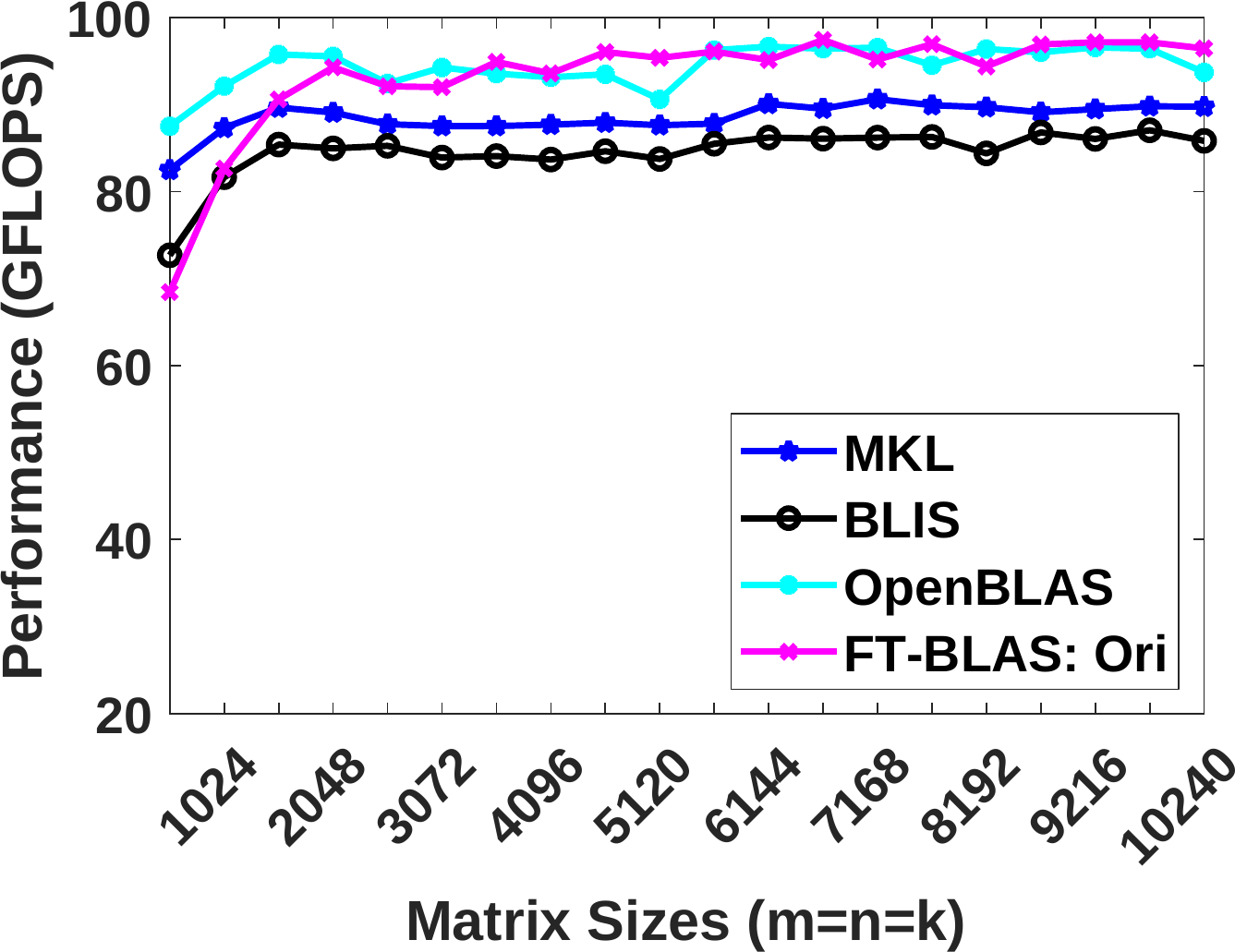}
}
\hspace{-3mm}
\vspace{-3mm}
\subfigure[DTRSM]
{
\includegraphics[width=0.21\textwidth]{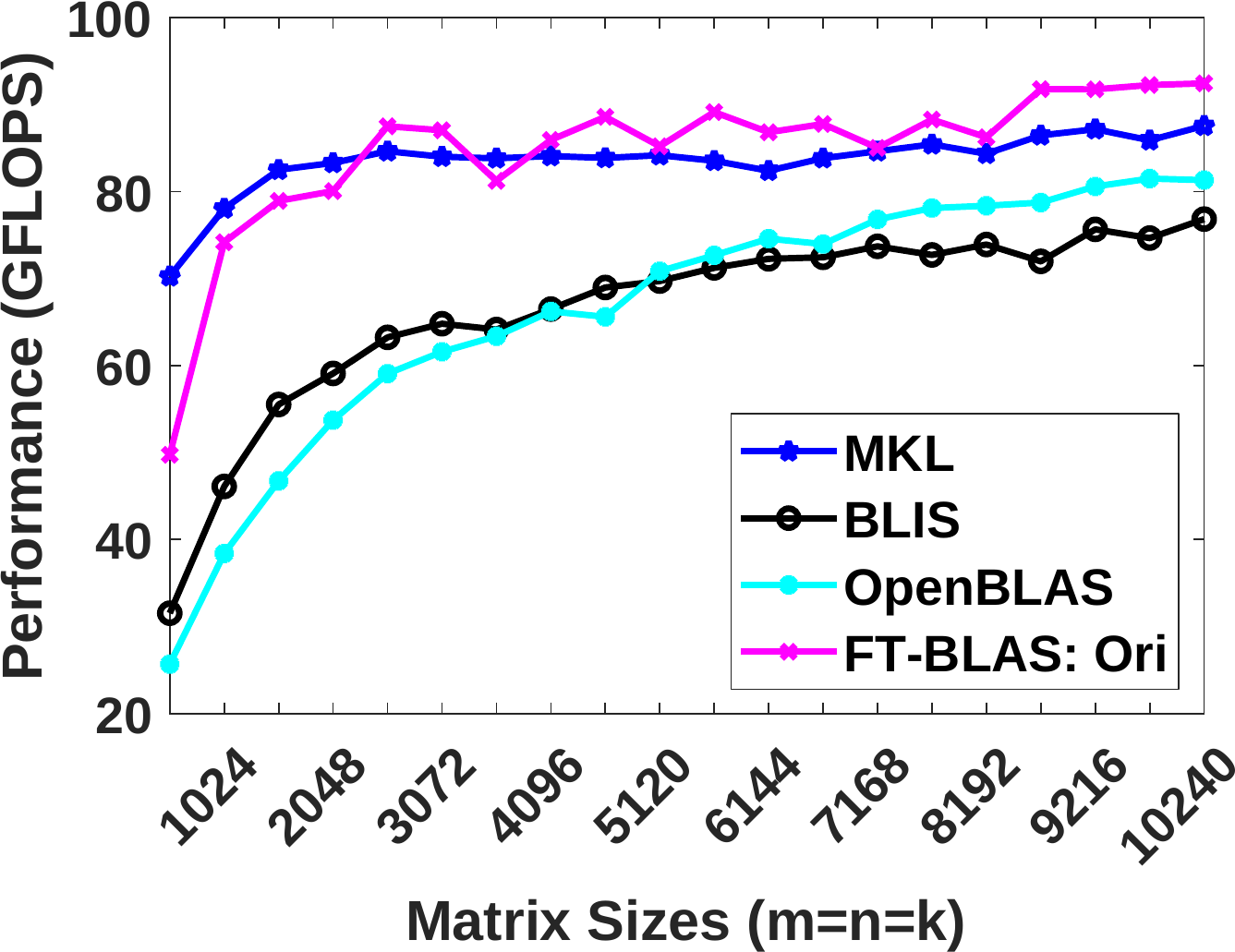}
}
\caption{Comparisons of selected Level-3 BLAS routines.}
\label{fig:other-level-3}
\vspace{-2mm}
\end{figure}

\subsubsection{Optimizing Level-3 BLAS}
Adopting the traditional cache blocking and packing scheme, our DGEMM performs similarly to OpenBLAS DGEMM. As seen in Figure \ref{fig:other-level-3}, both of these DGEMM implementations outperform MKL and BLIS by 7.29-11.75\%. For the Level-3 BLAS routine DTRSM, we provide a highly-optimized macro kernel to solve for the diagonal block and cast the majority of the computation to the near-optimal DGEMM. Because OpenBLAS and BLIS simply provide an unoptimized scalar implementation for the diagonal solver, our DTRSM outperforms OpenBLAS and BLIS by 22.19\% and 24.77\%, and surpasses MKL by 3.33\%.

\subsection{Performance of FT-BLAS with Fault Tolerance Capability}
Having achieved comparable or better performance than the current state-of-the-art BLAS libraries without fault tolerance, we now add on fault tolerance functionalities. For memory-bound Level-1 and Level-2 BLAS routines, we propose a novel DMR verification scheme based on the AVX-512 instruction set and then further reduce the overhead of fault tolerance to a negligible level via assembly optimization. For computing-bound Level-3 BLAS, we fuse the checksum calculations into the packing routines and assembly kernels to reduce data transfer between registers and memory. The results in this section were obtained with fault tolerant DMR and ABFT operating, but not under active fault injection---see subsection C for injection experiments.

\begin{figure}[ht] \centering
\subfigure[{Performance Optimization}]
{
\includegraphics[width=0.215\textwidth]{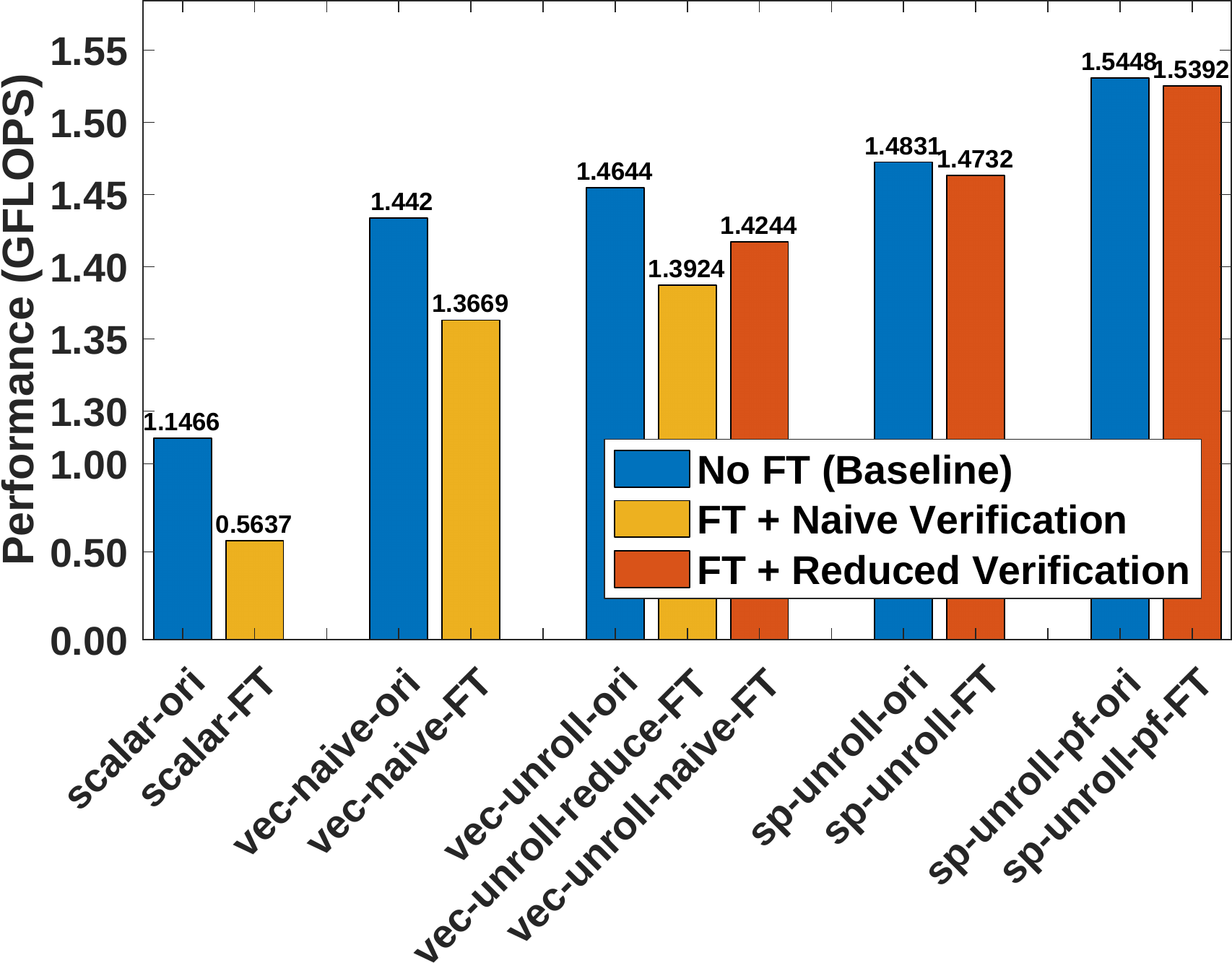}
}
\hspace{-3mm}
\vspace{-3mm}
\subfigure[Overhead Optimization]
{
\includegraphics[width=0.215\textwidth]{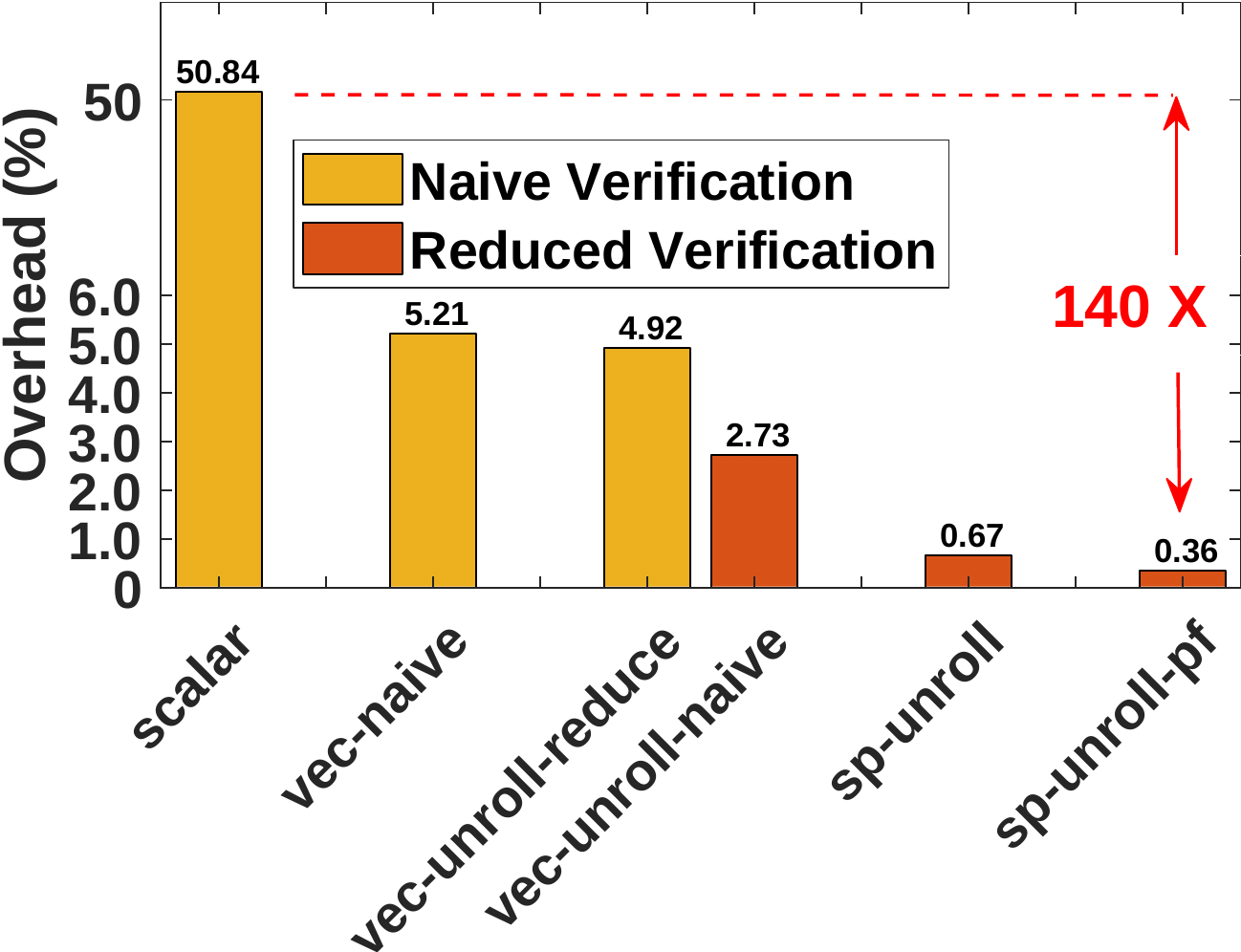}
}
\vspace{-1mm}
\caption{Optimizing DSCAL with/without FT.}
\label{fig:dscal-opt}
\vspace{-2mm}
\end{figure}
\subsubsection{Reducing DMR overhead for memory-bound routines}

Figure \ref{fig:dscal-opt} presents the performance and overhead of DSCAL with step-wise assembly level optimization. In each step, the assembly optimization described in Sections III and IV are applied to the FT version and its baseline, our non-FT version evaluated above. The performance of the most naive baseline, a scalar implementation, is 1.15 GFLOPs. Duplicating computing instructions and verifying correctness for this baseline halves the performance to 0.56 GFLOPS, bringing a 50.83\% overhead. A vectorized implementation based on AVX-512 instructions decreases overhead by 9.8-fold compared to the scalar duplication/verification scheme. A vectorized implementation with fault-tolerance capability increases performance to 1.36 GFLOPs, 2.42-fold of the scalar FT version. After this vectorization, simply unrolling the loop gains $1.55\%$ and $1.87\%$ improvement on the non-FT (vec-unroll-ori) and FT (vec-unroll-naive) versions respectively, while the overhead is now $4.9\%$. It is at this point that our non-FT version reaches OpenBLAS. Our novel verification scheme involving opmask registers improves the overhead to $2.7\%$. We then schedule instructions via heuristic software pipelining, improving the performance of the non-FT (sp-unroll-ori) and FT (sp-unroll-FT) implementations to 1.48 GFLOPs and 1.47 GFLOPs respectively. The overhead improves to $0.67\%$ in this step. We add prefetch instructions as a final step, and the overhead settles at $0.36\%$.

\begin{figure}[ht] \centering
\vspace{-2mm}
\subfigure[{Performance of FT-DGEMM}]
{
\includegraphics[width=0.215\textwidth]{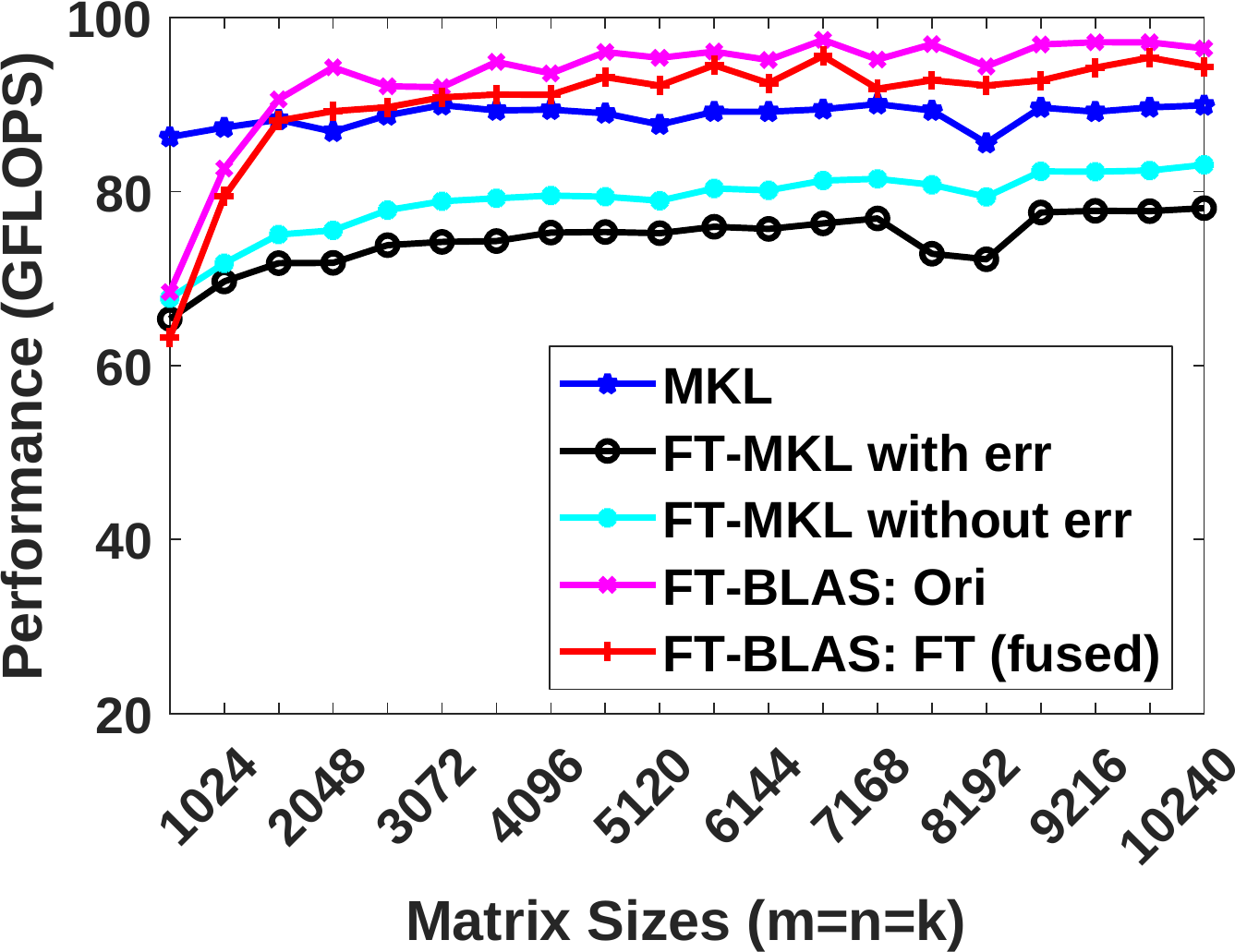}
}
\hspace{-3mm}
\vspace{-3mm}
\subfigure[Linking to different libraries]
{
\includegraphics[width=0.21\textwidth]{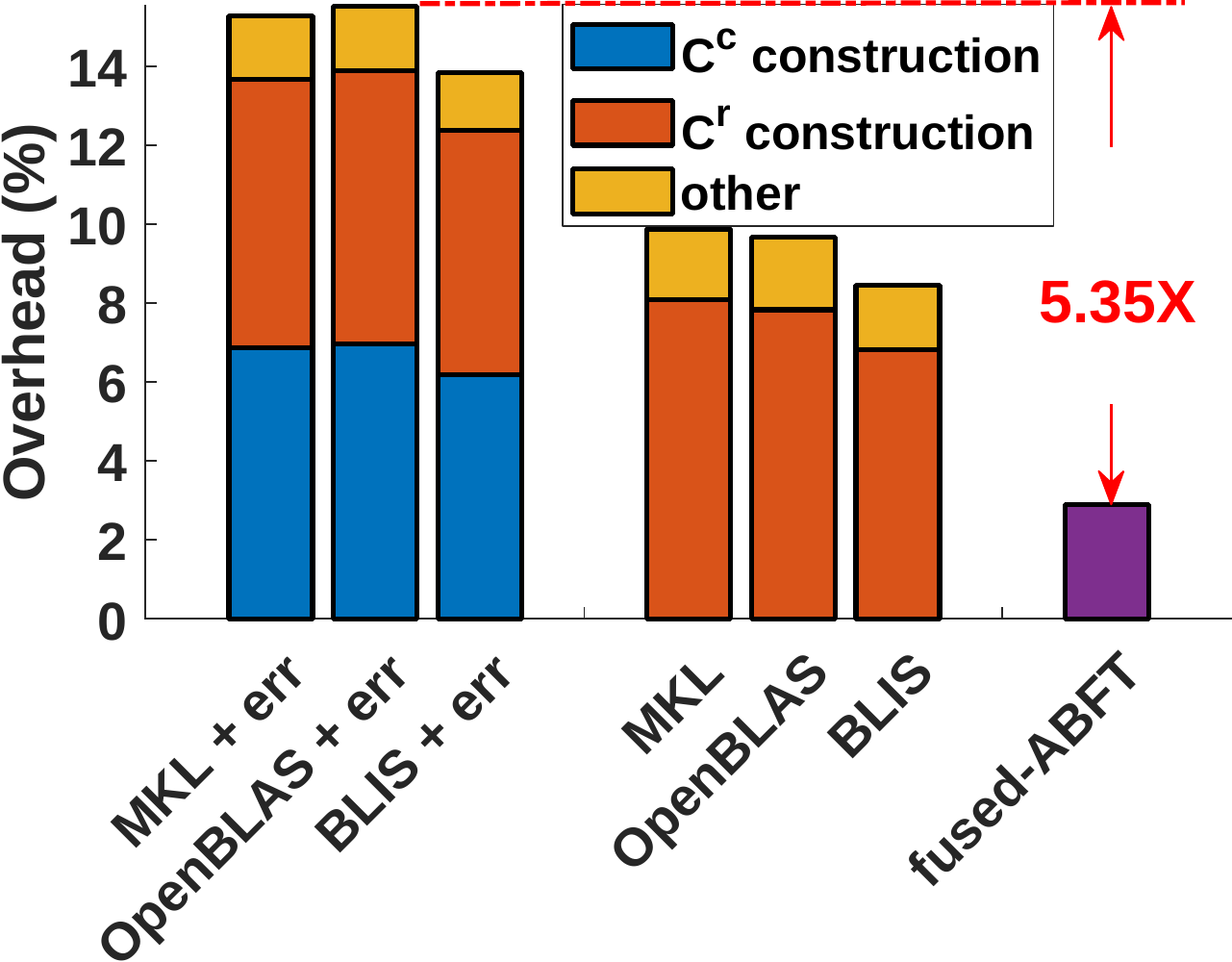}
}
\vspace{-1mm}
\caption{Optimizing DGEMM with FT.}
\label{fig:dgemm-ft-opt}
\vspace{-6mm}
\end{figure}

\subsubsection{Reducing ABFT overhead for computing-bound routines}

Figure \ref{fig:dgemm-ft-opt} (a) presents the performance of two methods of implementing ABFT for GEMM: building upon MKL (FT-MKL) and fusing into the GEMM routine (FT-BLAS: FT fused). FT-MKL under error injection leads to 15\% overhead compared with baseline MKL. When there is no error injected, we no longer compute and verify the checksum $C^r$ so the overhead decreases to 9\%. In contrast, the fused implementation (2.9\% overhead) of ABFT does not generate an extra cost when encountering errors because its reference checksum computation is fused into the assembly computing kernel and is computed regardless of whether an error is detected. As shown in Figure \ref{fig:dgemm-ft-opt} (b), the overhead of building ABFT on a third-party library slightly varies when linking to different libraries but the trend is clear: reference checksum construction generates the majority of the ABFT overhead, which is eliminated by the fusing strategy. The overhead can be up to 5.35-fold that of fusing ABFT into DGEMM. Our overhead is also lower than Smith et al's work in 2015 \cite{smith2015toward}, where checkpoint/rollback recovery is used to tolerate errors. Their checkpoint/rollback recovery has a wider error coverage, but the overhead is ``in the range of 10\%"\cite{smith2015toward}.

\begin{figure}[] 
\centering
\subfigure[FT-DSCAL]
{
\includegraphics[width=0.21\textwidth]{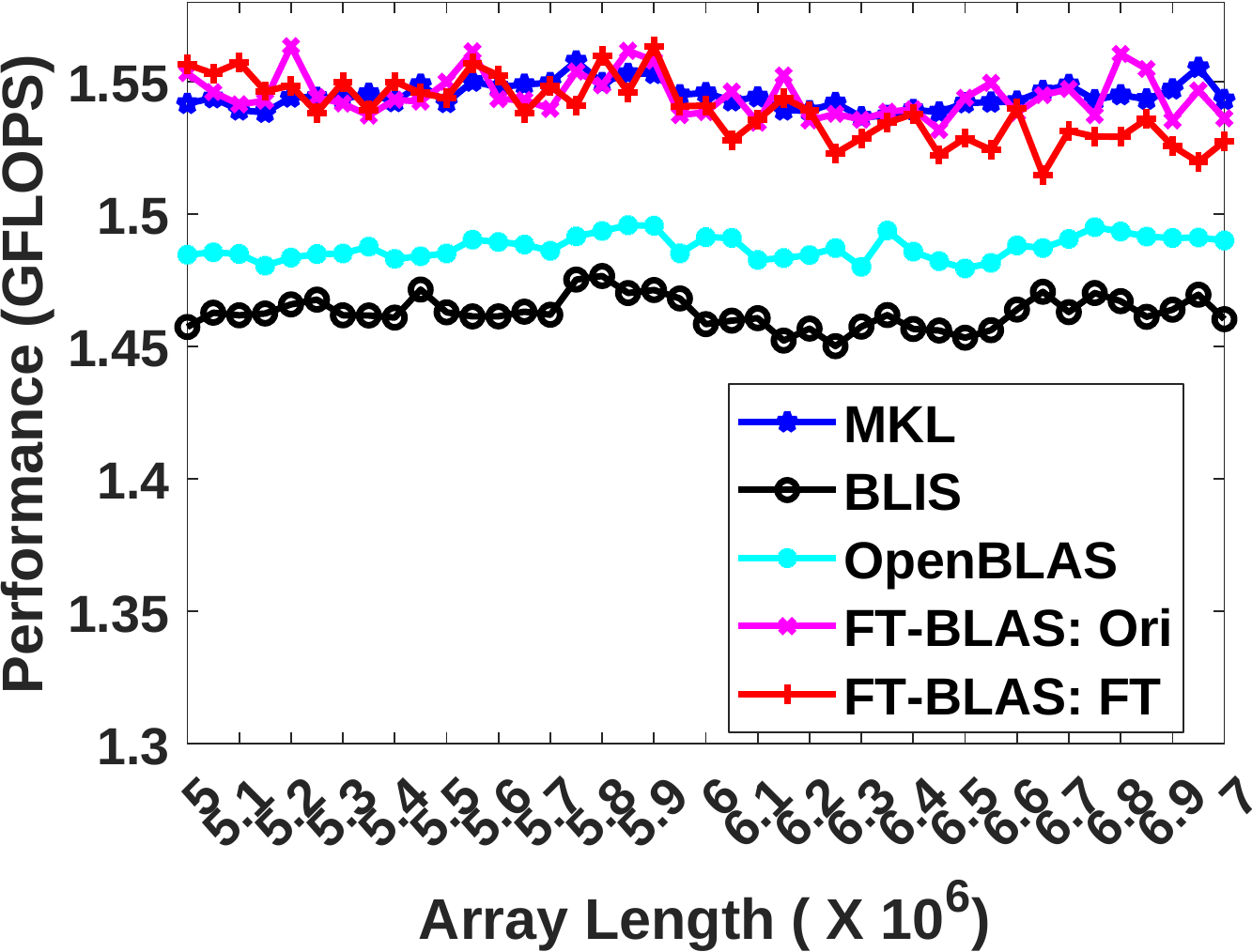}
}
\vspace{-1mm}
\hspace{-3mm}
\subfigure[FT-DNRM2]
{
\includegraphics[width=0.21\textwidth]{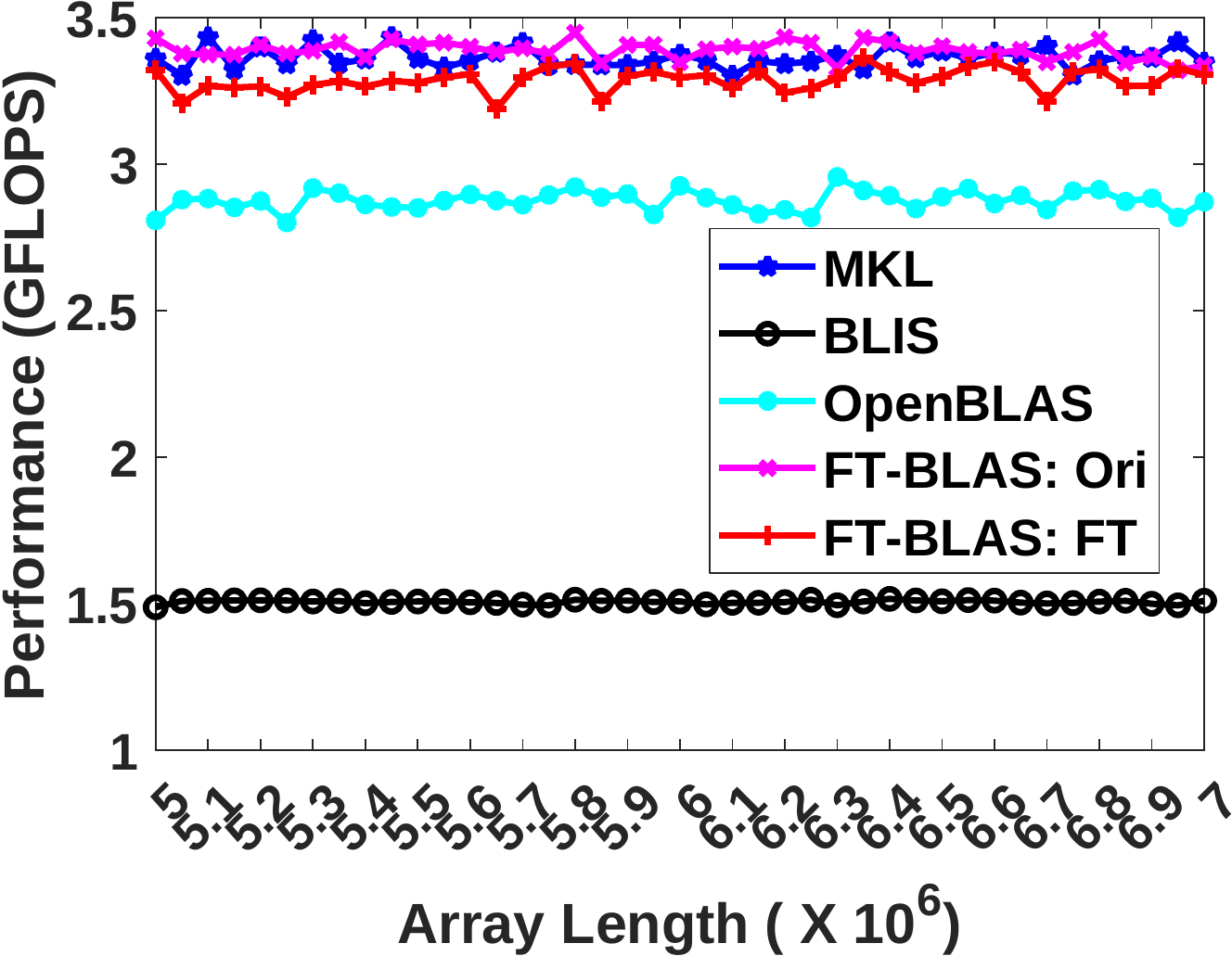}
}
\vspace{-1mm}
\hspace{-3mm}
\subfigure[FT-DGEMV]
{
\includegraphics[width=0.21\textwidth]{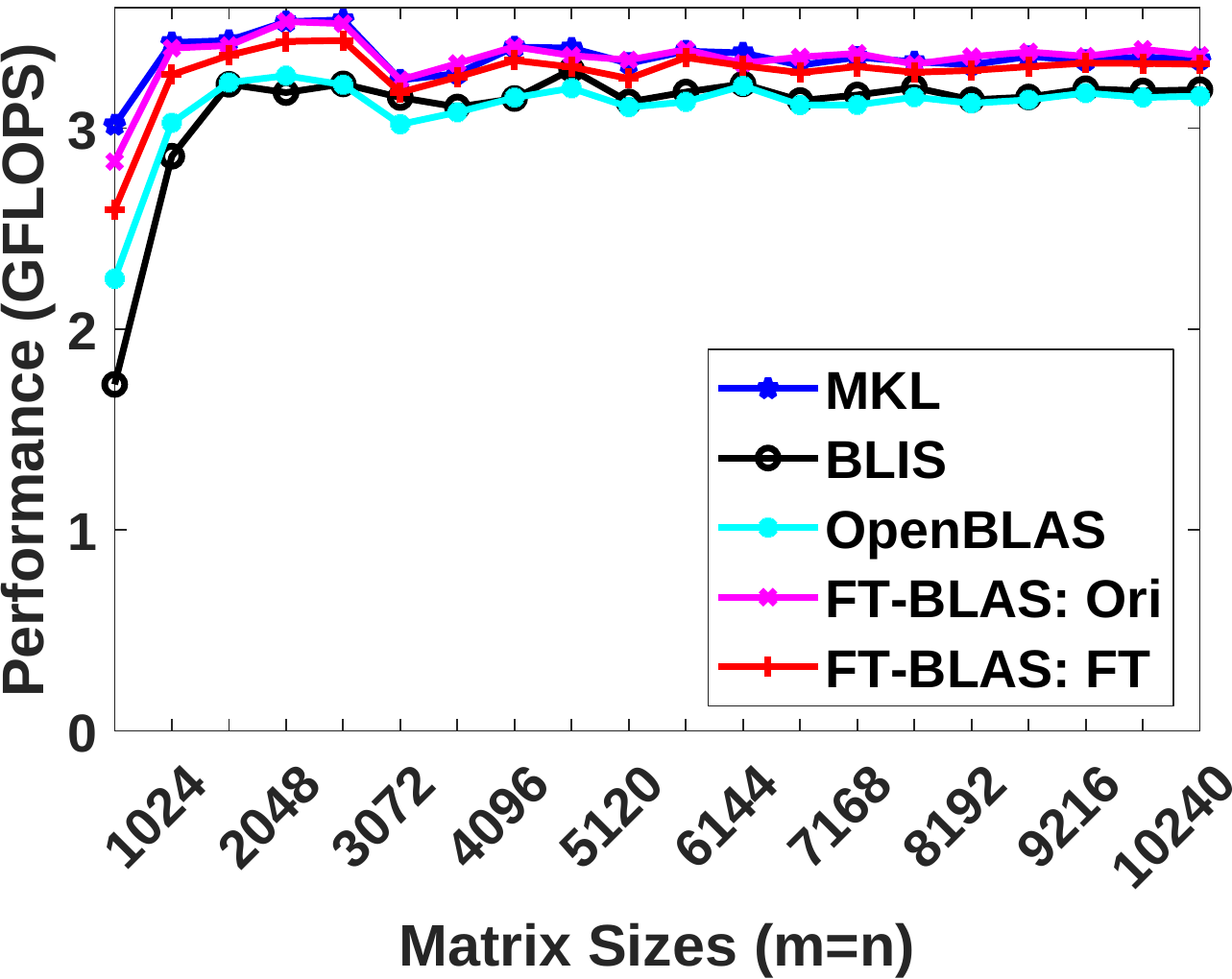}
}
\vspace{-1mm}
\hspace{-3mm}
\subfigure[FT-DTRSV]
{
\includegraphics[width=0.21\textwidth]{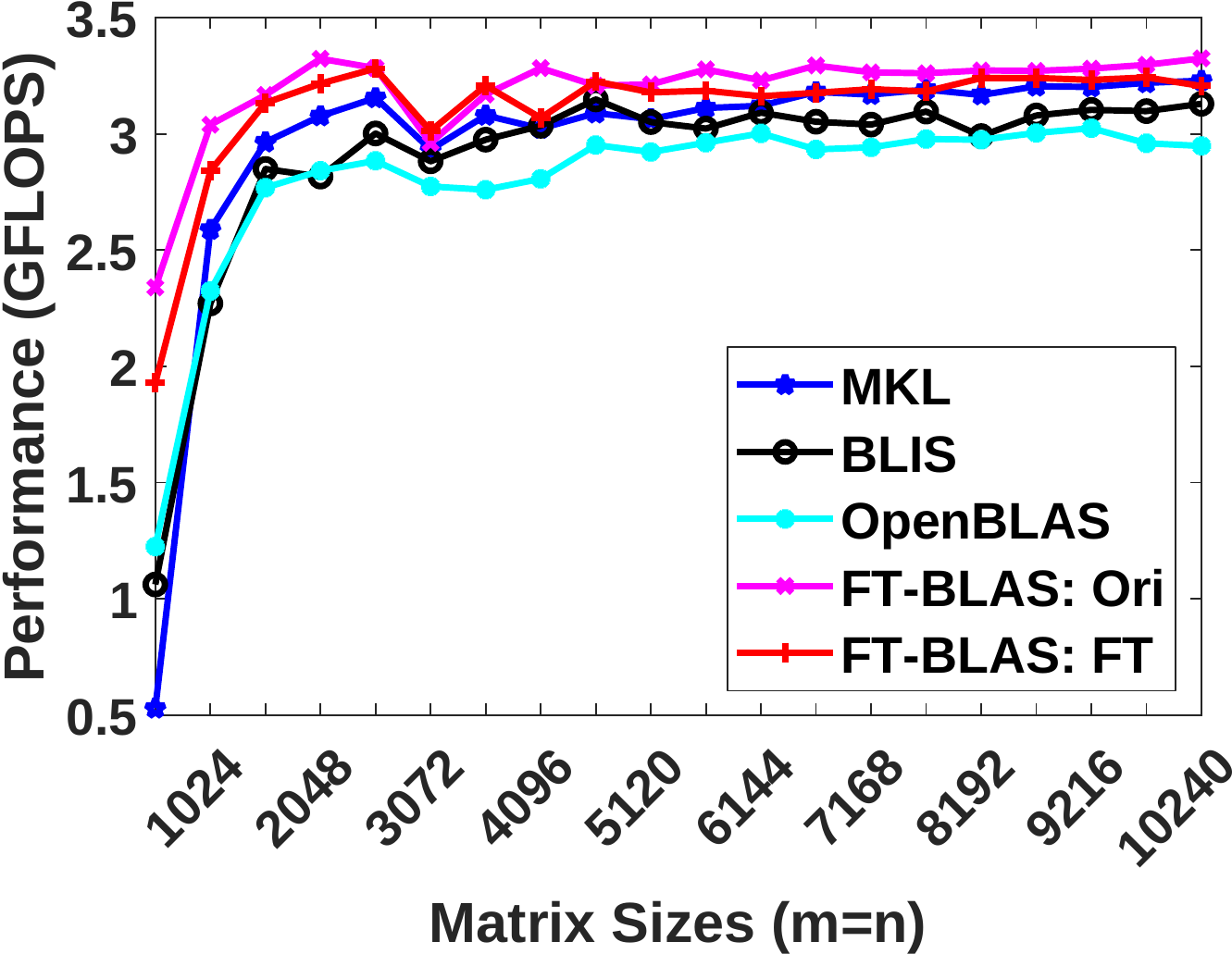}
}
\vspace{-1mm}
\hspace{-3mm}
\subfigure[FT-DGEMM]
{
\includegraphics[width=0.21\textwidth]{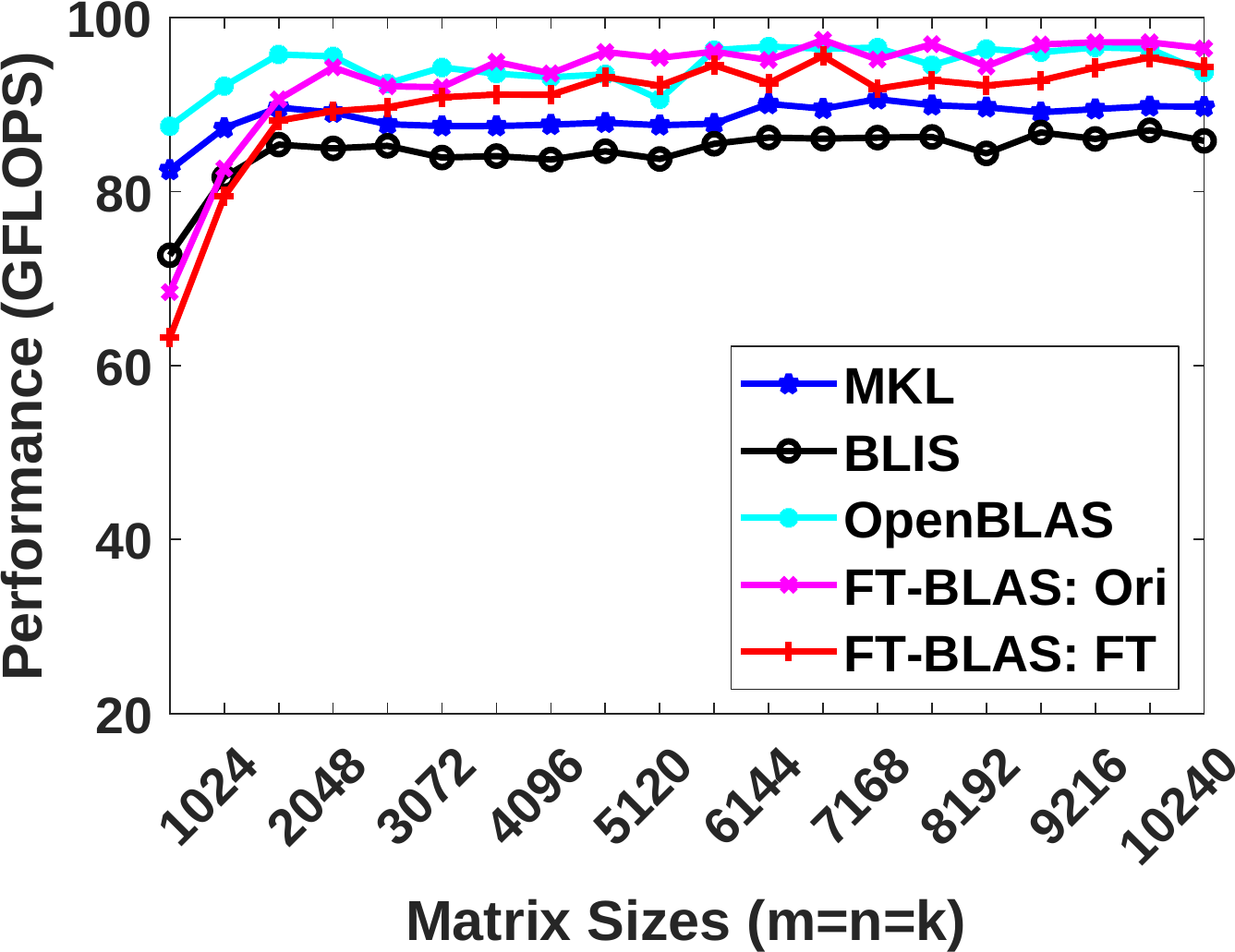}
}
\vspace{-1mm}
\hspace{-3mm}
\subfigure[FT-DSYMM]
{
\includegraphics[width=0.21\textwidth]{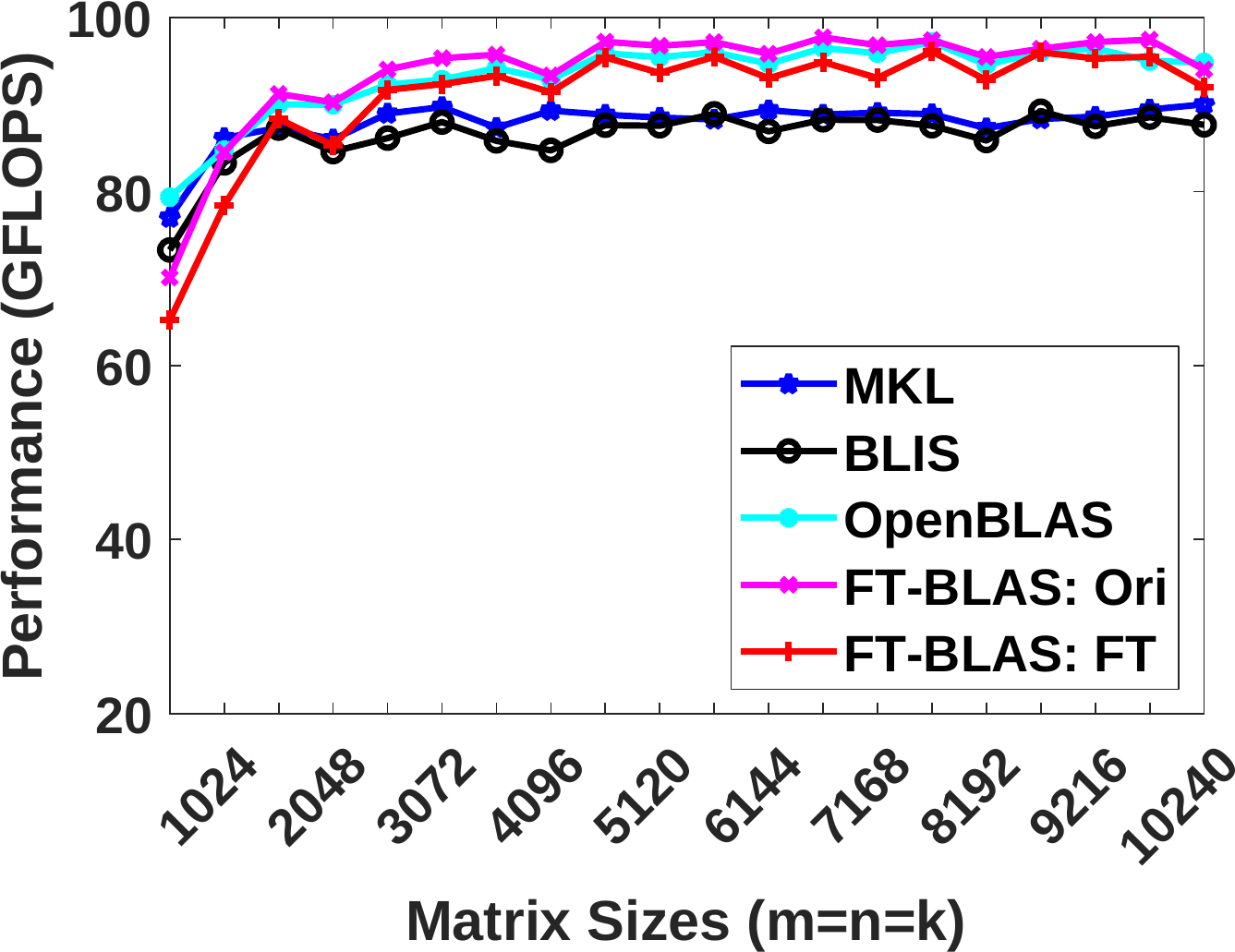}
}
\vspace{-1mm}
\hspace{-3mm}
\subfigure[FT-DTRMM]
{
\includegraphics[width=0.21\textwidth]{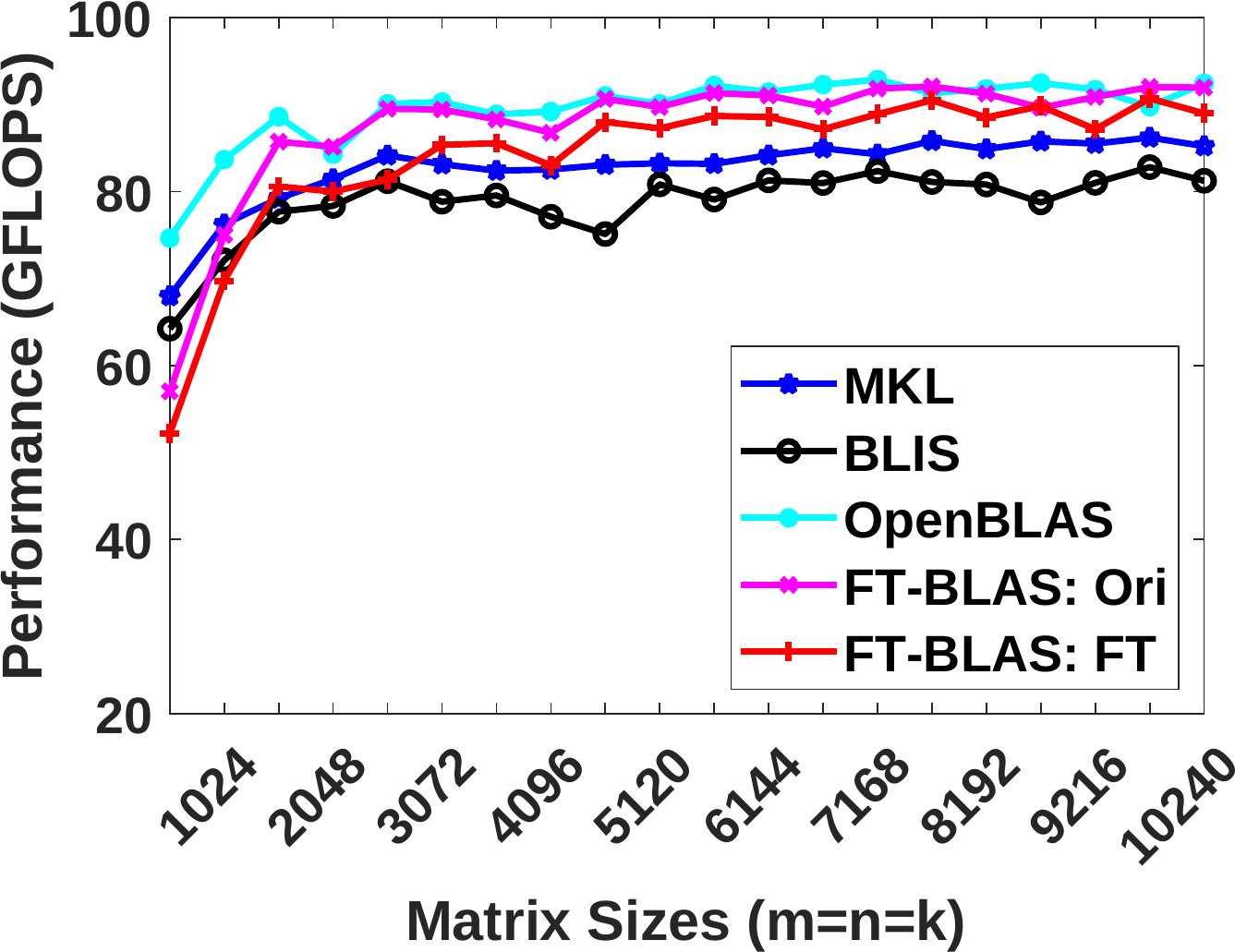}
}
\vspace{-1mm}
\hspace{-3mm}
\subfigure[FT-DTRSM]
{
\includegraphics[width=0.21\textwidth]{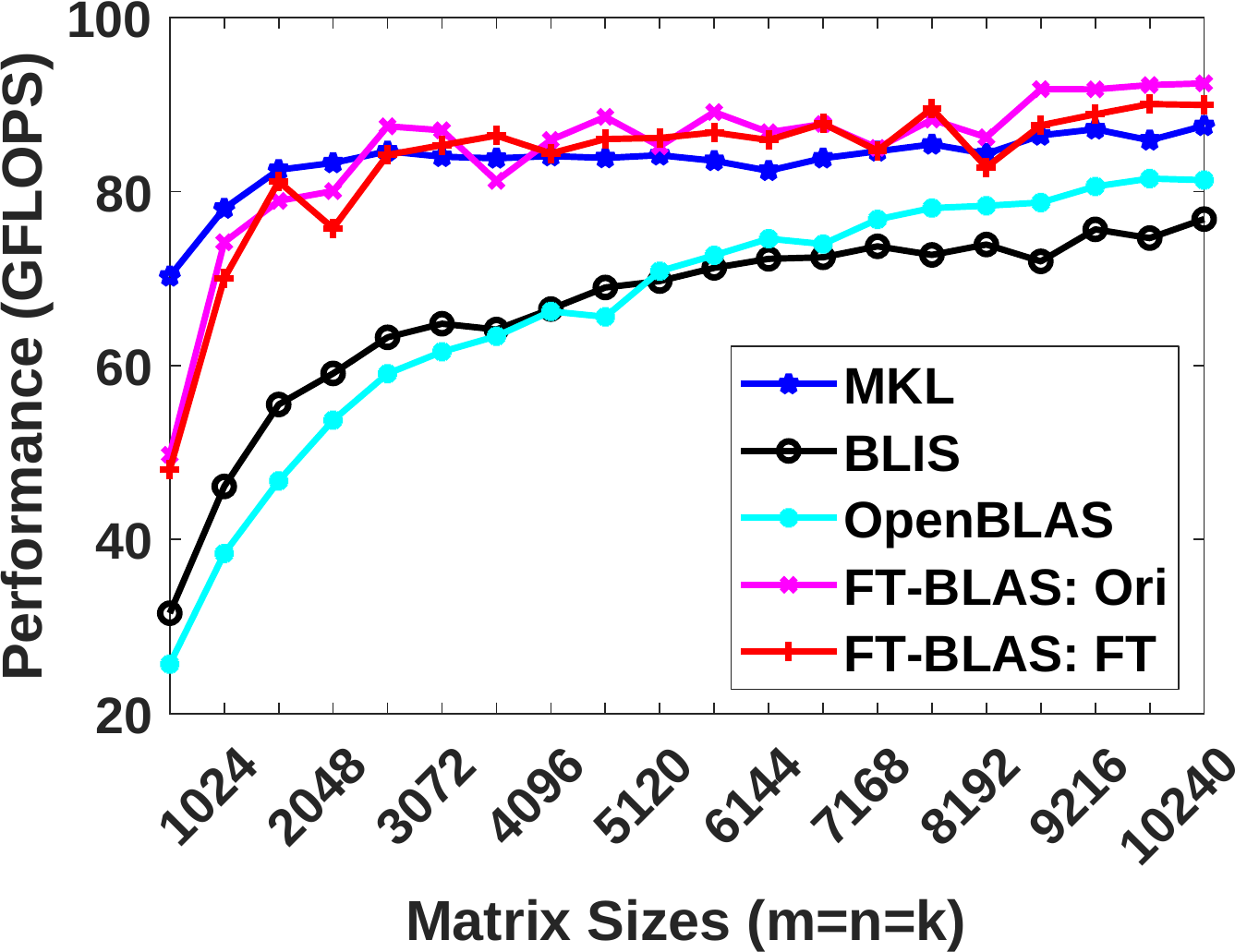}
}
\caption{Comparisons of selected BLAS routines with FT.}
\label{fig:ft-blas}
\vspace{-4mm}
\end{figure}

\subsubsection{Generalizing to other routines}
Figure \ref{fig:ft-blas} compares the performance of FT-BLAS with FT capability (FT-BLAS: FT) against its baseline: our implementation without FT capability (FT-BLAS: Ori) and reference BLAS libraries on eight routines of all three levels of BLAS. The DMR-based FT implementations for the Level-1 and Level-2 BLAS routines (DSCAL, DNRM2, DGEMV, DTRSV) generate 0.34\%-3.10\% overhead over baseline. For the Level-3 BLAS routines, DGEMM, DSYMM, DTRMM and DTRSM, our strategy to fuse memory-bound ABFT operations with matrix computation generates overhead ranging from 1.62\% to 2.94\% on average. Our implementation strategy for DSYMM in both FT-BLAS: Ori and FT-BLAS: FT is similar to the DGEMM scheme, with moderate modification to the packing routines. For DTRMM, we use the same strategy with some additional modifications to the computing kernel, similar to the methods in \cite{goto2008high}. With these negligible overheads added to an already high-performance baseline, our FT-BLAS with FT capability remains comparable to or faster than the reference libraries.

\subsection{Error Injection Experiments}
We validate the effectiveness of our fault-tolerance scheme by injecting multiple computing errors into each of our computing kernels and verifying our final computation results against MKL. External error injection tools often significantly slow down the native program\cite{luk2005pin, oliveira2017carol, guan2014f}, therefore, we inject errors from a source code level to minimize the performance impact on native programs.

We inject 20 errors into each routine. The length of the injection interval $k$ is determined based on the number of errors to inject, that is, we inject one error every $k$ iterations. For ABFT-protected Level-3 BLAS routines, the error injection is straightforward because we can directly operate in C code. An element of matrix $C$ is randomly selected for modification when an injection point is reached. This injected error will lead to a difference in the checksum relationship, and the erroneous element and error magnitude will be computed accordingly. This detected error is then corrected by subtracting the error magnitude from the erroneous position. For DMR-protected Level-1 and Level-2 BLAS routines, the injection is more complicated since the loop body is implemented purely using assembly codes. Therefore, providing an assembly-level error injection mechanism becomes necessary. Once the program reaches an injection point, we redirect the control flow to a faulty loop body to generate an error. This generated error is then detected via comparison with the computed results of the duplicated instruction. After the error is detected, a recovery procedure is activated to recompute the corrupted iteration immediately. In all cases we validate the correctness our final computations by comparing with MKL to ensure all injected errors were truly corrected.

\begin{figure}[ht] 
\centering
\subfigure[DGEMV with error injection]
{
\includegraphics[width=0.205\textwidth]{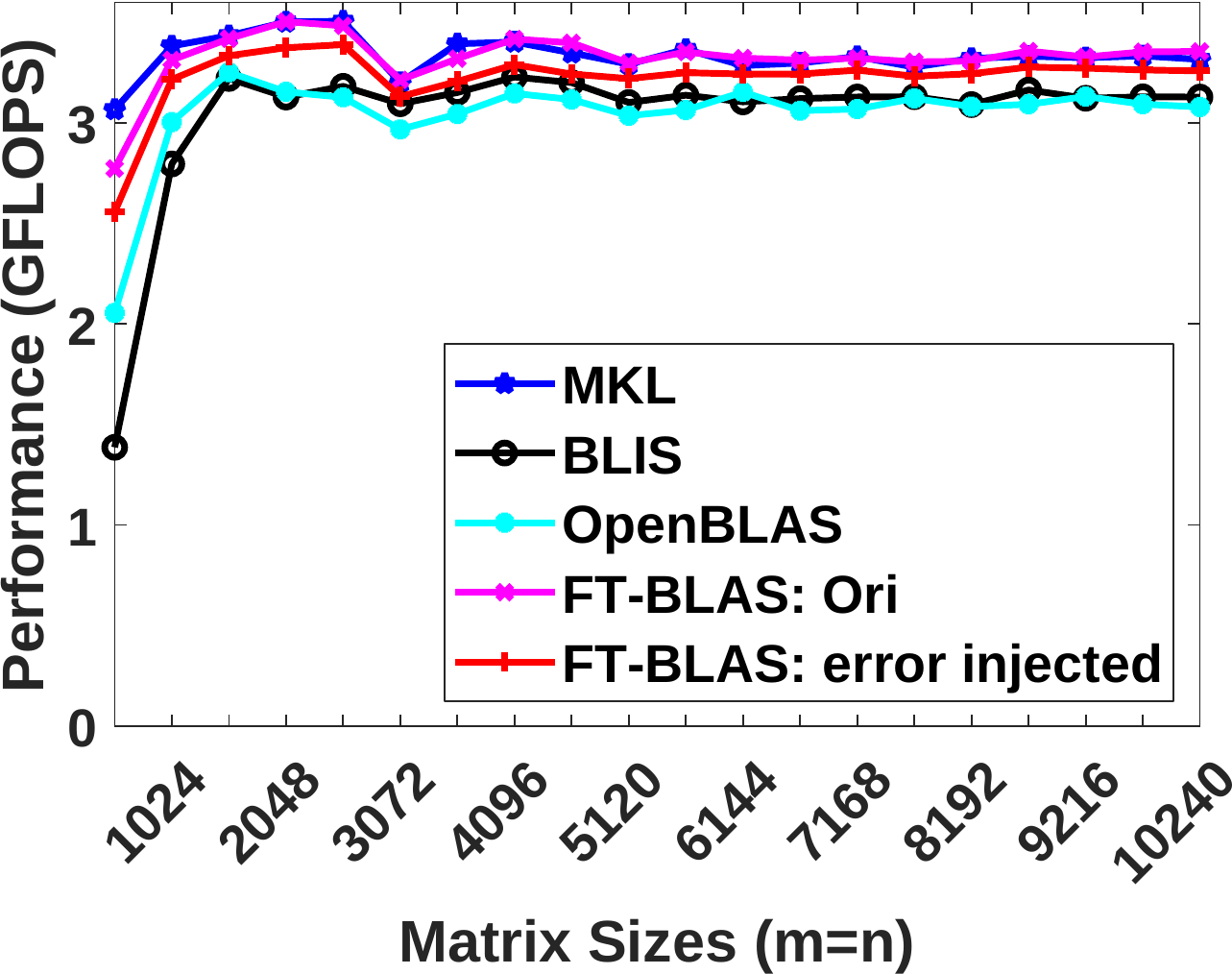}
}
\hspace{-3mm}
\vspace{-2mm}
\subfigure[DTRSV with error injection]
{
\includegraphics[width=0.21\textwidth]{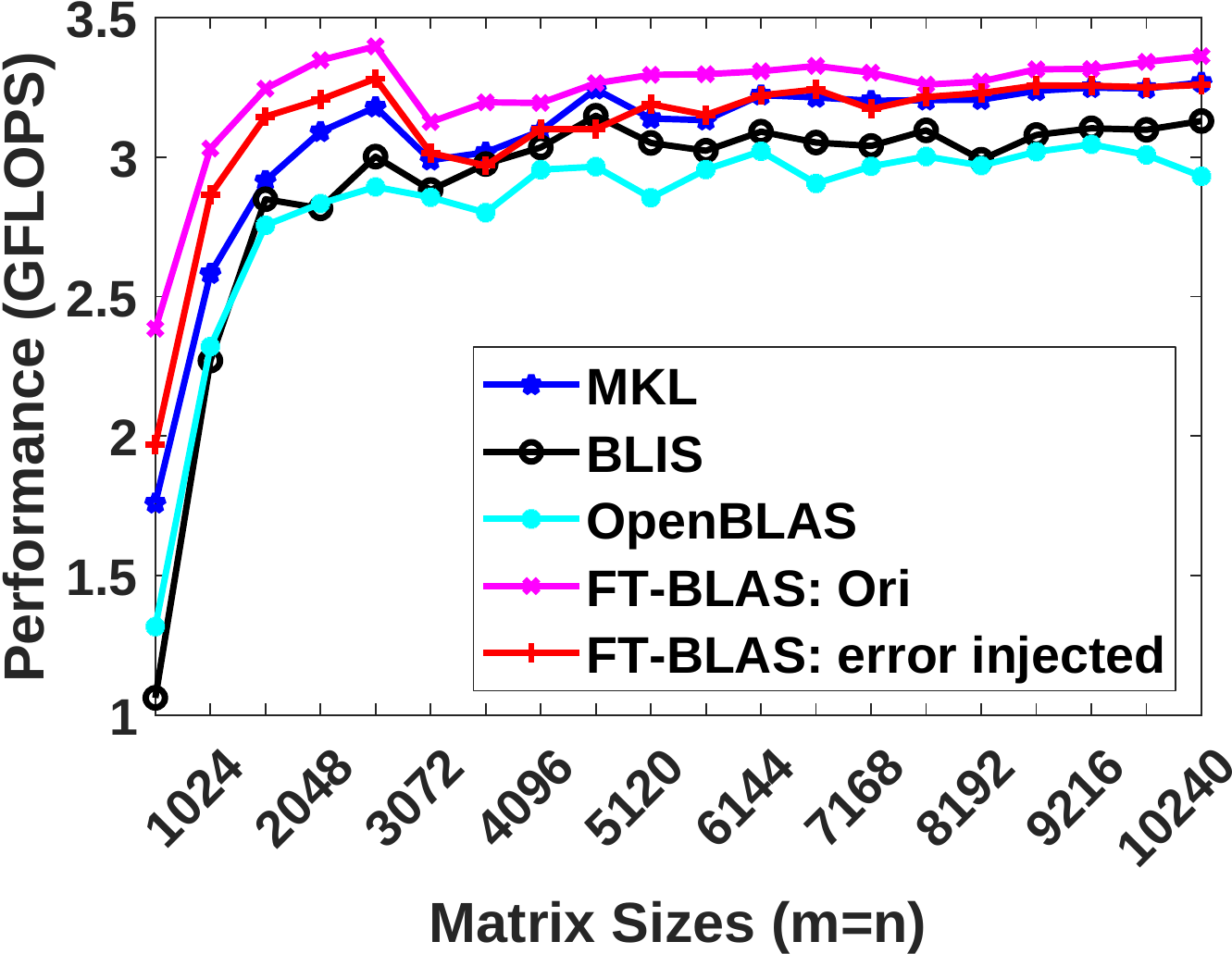}
}
\hspace{-3mm}
\vspace{-2mm}
\subfigure[DGEMM with error injection]
{
\includegraphics[width=0.205\textwidth]{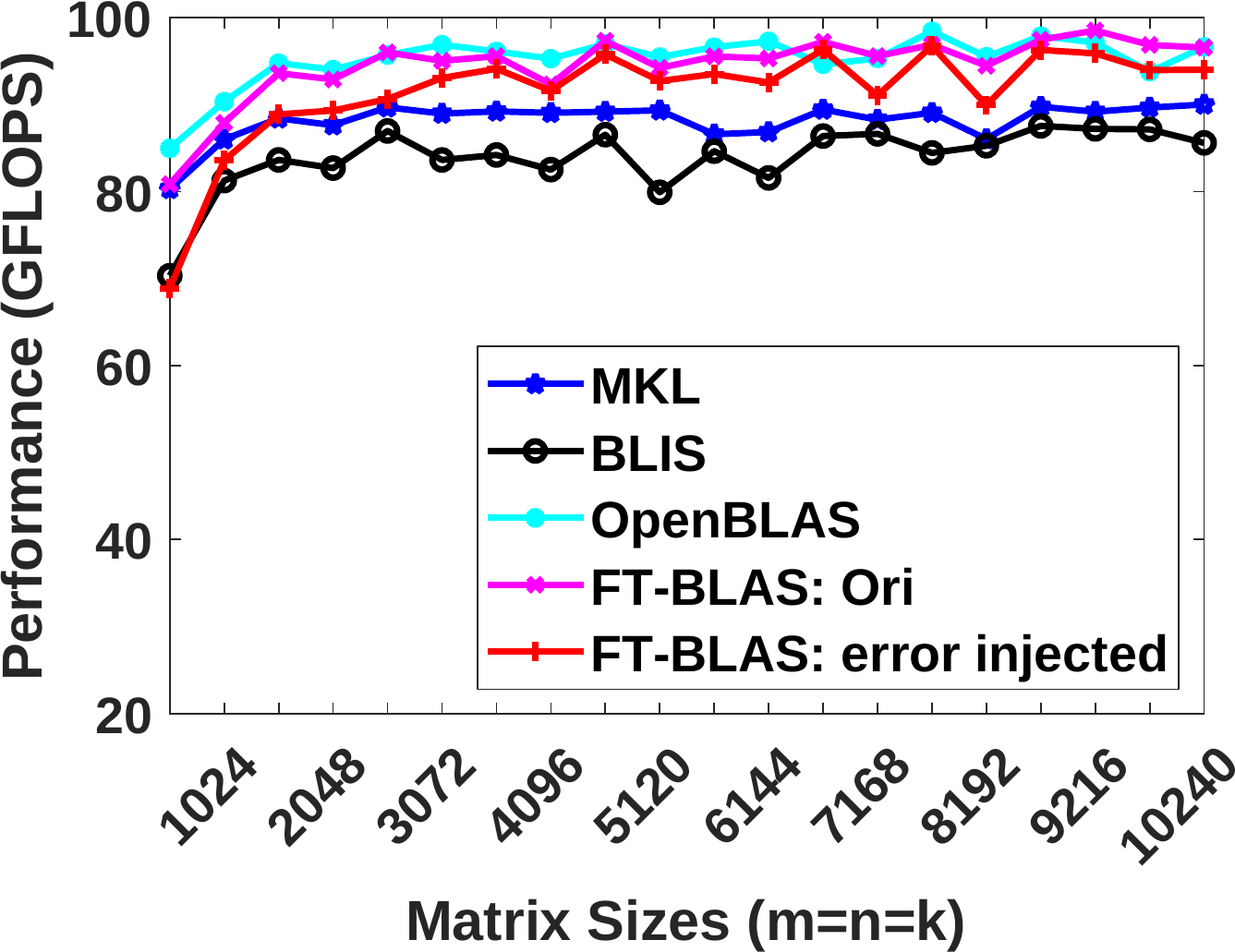}
}
\hspace{-3mm}
\subfigure[DTRSM with error injection]
{
\includegraphics[width=0.21\textwidth]{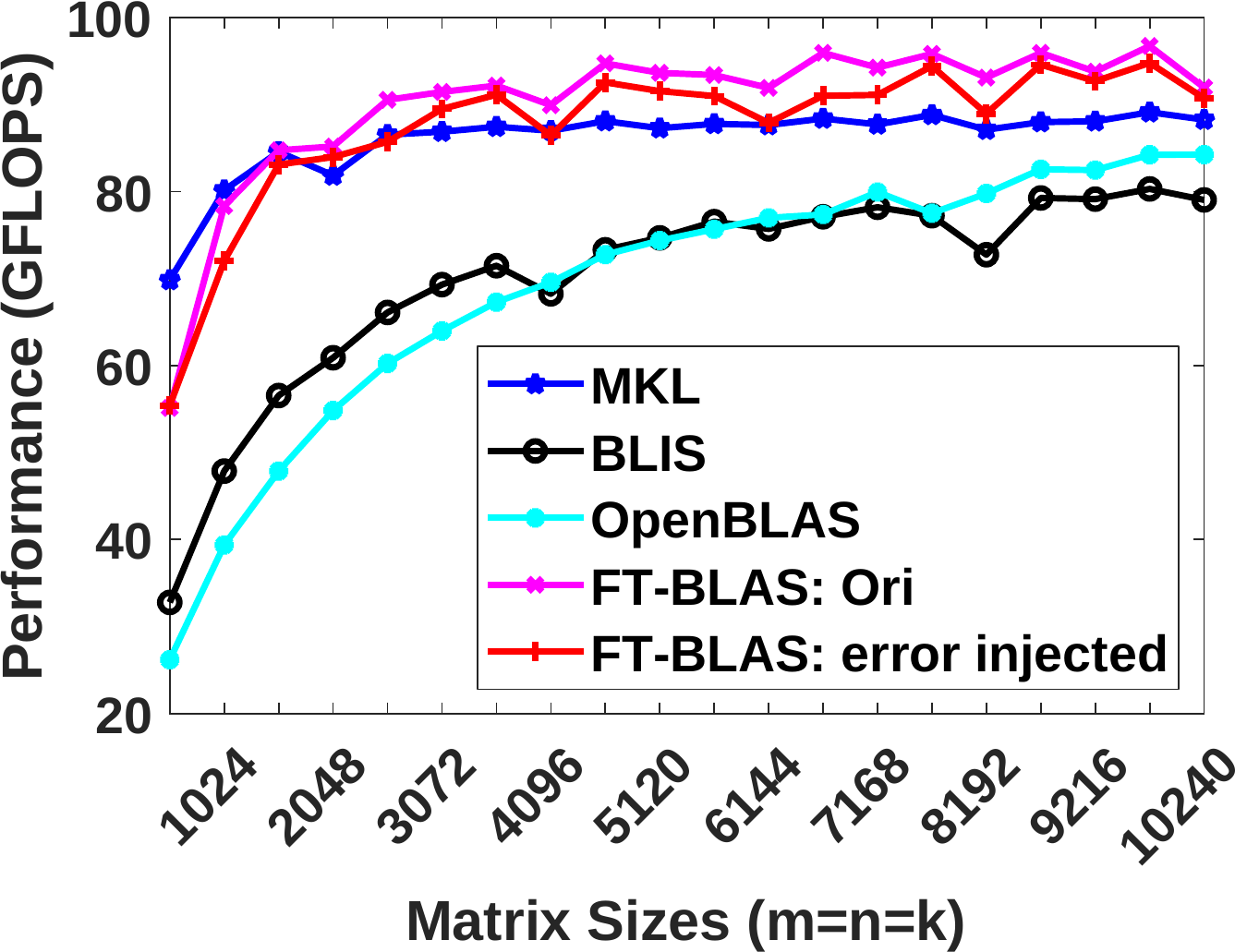}
}
\hspace{-3mm}
\caption{Performance under error injection.}
\label{fig:err-inj-skl}
\vspace{-3mm}
\end{figure}

Figure \ref{fig:err-inj-skl} compares the performance of four routines under error injection. For both DMR-protected (DGEMV, DTRSV) and ABFT-protected (DGEMM, DTRSM) routines, we maintain negligible (2.47\%-3.22\%) overhead and the overall performance under error injection remains comparable or faster than reference libraries. In particular, our DTRSM outperforms OpenBLAS, BLIS, and MKL by 21.70\%, 22.14\%, and 3.50\% even under error injection. Experimental results confirm that our protection schemes do not require significant extra overhead to correct errors. This is because our correction methods---either to recompute the corrupted iteration or to subtract an error magnitude from the incorrect position---generate only a few ALU computations instead of expensive memory accesses. 

\begin{figure}[ht] 
\centering
\subfigure[DTRSV with error injection]
{
\includegraphics[width=0.205\textwidth]{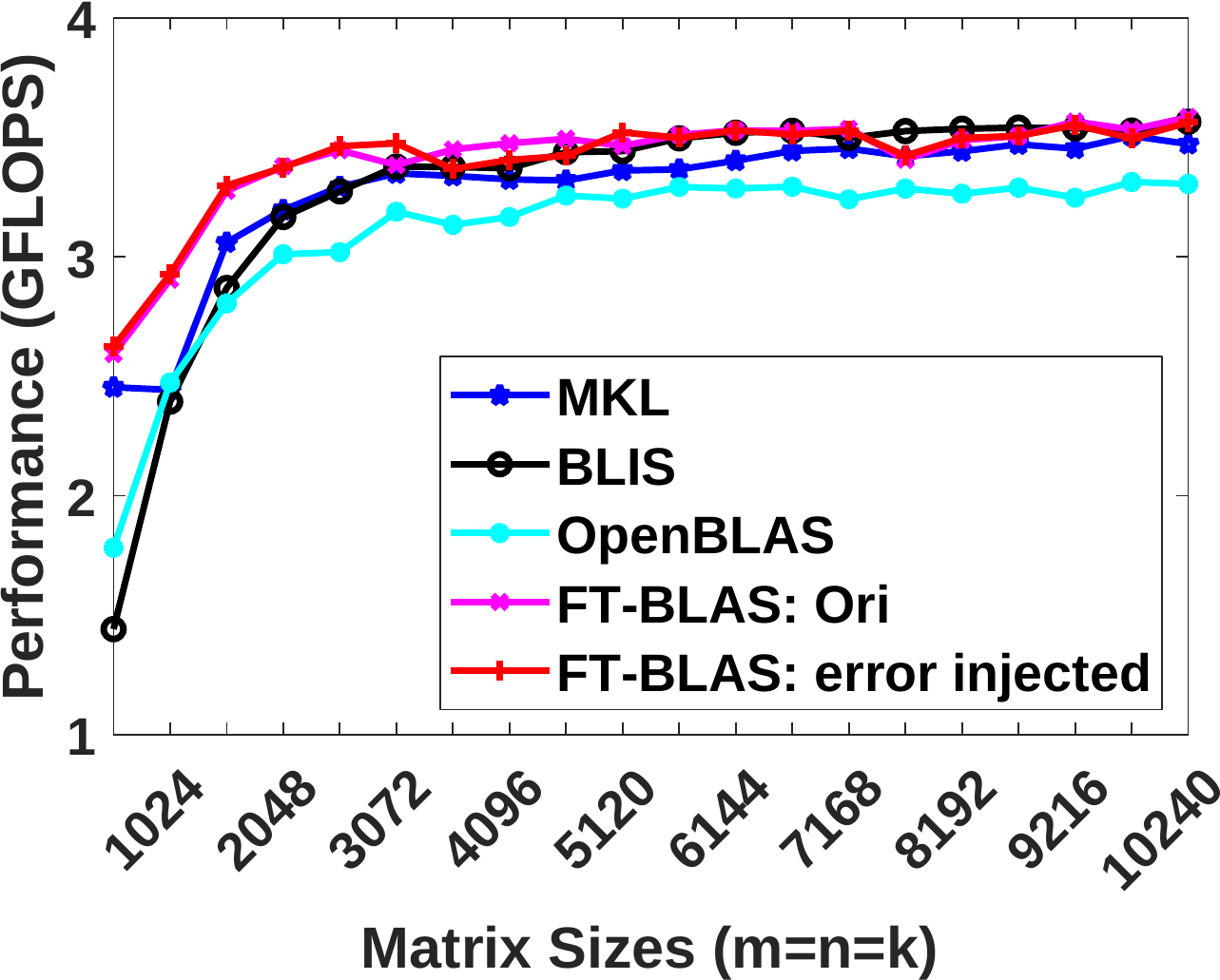}
}
\hspace{-3mm}
\vspace{-2mm}
\subfigure[DTRSM with error injection]
{
\includegraphics[width=0.21\textwidth]{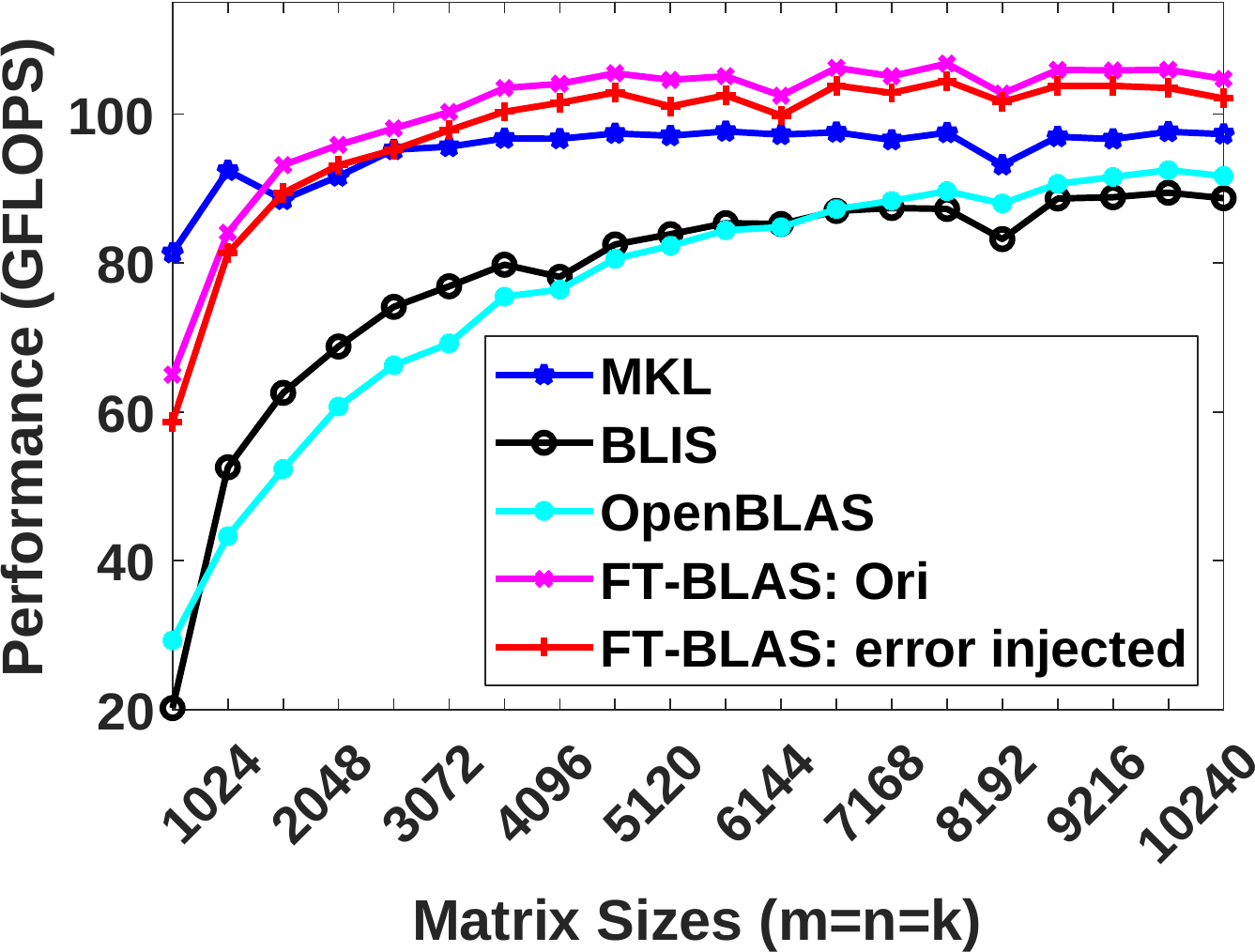}
}
\hspace{-3mm}
\vspace{-2mm}
\caption{Performance under error injection on Cascade Lake.}
\label{fig:err-inj-csl}
\end{figure}

We further test FT-BLAS under error injection using another processor, the Intel Cascade Lake W-2255. As shown in Figure \ref{fig:err-inj-csl}, our protection scheme is as lightweight as it was on the Skylake processor, and is still able to surpass open-source OpenBLAS and BLIS by 22.89\% and 21.56\% and the closed-source MKL by 4.98\% even while tolerating 20 injected errors. The execution time of DTRSM and DTRSV for $512^2$ to $10240^2$ matrices ranges from 2 ms to 20 seconds. Therefore, injecting 20 errors into these two routines is equivalent to injecting 1 to 10,000 errors per second. Hence, FT-BLAS is able to tolerate up to thousands of errors per second with comparable and sometimes faster performance than state-of-the-art BLAS libraries---and none of them can tolerate soft errors. Error injection results for other routines are similar, but due to page limits these results are skipped here.

\section{Conclusions}
We present a fault-tolerant BLAS implementation that is not only capable of tolerating soft errors, but also achieves comparable or superior performance over the current state-of-the-art libraries, OpenBLAS, BLIS, and Intel MKL. Future work will focus on extending FT-BLAS to more architectures with parallel support and open-sourcing the code.
 
\section{Acknowledgements}
This work was supported by National Science Foundation (NSF) CAREER Award 1305624 and University of
California, Riverside Academic Senate Committee on Research (CoR) Grant. We thank the anonymous reviewers for their insightful comments.

\bibliographystyle{ACM-Reference-Format}
\bibliography{sample-base}
\end{document}